\RequirePackage[utf8]{inputenc}
\RequirePackage[T1]{fontenc}

\documentclass[journal]{IEEEtran}

\usepackage{cite}

\ifCLASSINFOpdf
  \usepackage[pdftex]{graphicx}
  \DeclareGraphicsExtensions{.pdf,.jpeg,.png}
\else
  \usepackage[dvips]{graphicx}
  \graphicspath{{../eps/}}
  \DeclareGraphicsExtensions{.eps}
\fi

\usepackage{amsmath}
\usepackage{array}

\ifCLASSOPTIONcompsoc
  \usepackage[caption=false,font=normalsize,labelfont=sf,textfont=sf]{subfig}
\else
  \usepackage[caption=false,font=footnotesize]{subfig}
\fi

\usepackage{url}

\usepackage{siunitx} 

\usepackage{amsfonts,amssymb} 
\usepackage{tabularx} 
\usepackage{textcomp,gensymb} 

\newcommand{\Mw}[1]{M\textsubscript{w} {#1}}

\usepackage{xr} 
\makeatletter
\newcommand*{\addFileDependency}[1]{
  \typeout{(#1)}
  \@addtofilelist{#1}
  \IfFileExists{#1}{}{\typeout{No file #1.}}
}
\makeatother

\newcommand*{\myexternaldocument}[1]{%
    \externaldocument[ext-]{#1}
    \addFileDependency{#1.tex}%
    \addFileDependency{#1.aux}%
}
\myexternaldocument{supp_mat}

\begin{document}

\title{Deep Learning-based Damage Mapping with InSAR Coherence Time Series}
%
%
%

\author{Oliver~L.~Stephenson,
        Tobias~Köhne,
        Eric~Zhan,
        Brent~E.~Cahill,
        Sang-Ho~Yun,
        Zachary~E.~Ross
        and~Mark~Simons \thanks{Part of this research was supported by the National Aeronautics and Space Administration Applied Sciences Disasters Program and performed at the Jet Propulsion Laboratory, California Institute of Technology.}\thanks{O. L. Stephenson (corresponding author, olstephe@caltech.edu), T. Köhne, Z. E. Ross and M. Simons are with the Seismological Laboratory, Division of Geological and Planetary Sciences, California Institute of Technology, Pasadena, CA 91125, USA.}\thanks{E. Zhan is with the Department of Computing and Mathematical Sciences, Division of Engineering and Applied Science, California Institute of Technology, Pasadena, CA 91125, USA.}\thanks{B. E. Cahill is with REMY Biosciences, 17802 Sky Park Circle, Suite 100, Irvine, CA 92614, USA.}\thanks{S.-H. Yun is with the Jet Propulsion Laboratory, California Institute of Technology, Pasadena, CA 91109, USA.}}

%
%

\markboth{PAPER ACCEPTED FOR PUBLICATION IN IEEE TRANSACTIONS ON GEOSCIENCE AND REMOTE SENSING}{}%
%

\IEEEoverridecommandlockouts
\IEEEpubid{\begin{minipage}{\textwidth}\ \\[12pt] \copyright~2021 IEEE. Personal use of this material is permitted.  Permission from IEEE must be obtained for all other uses, in any current or future media, including reprinting/republishing this material for advertising or promotional purposes, creating new collective works, for resale or redistribution to servers or lists, or reuse of any copyrighted component of this work in other works. \end{minipage}}





\maketitle


\begin{abstract}
Satellite remote sensing is playing an increasing role in the rapid mapping of damage after natural disasters. In particular, synthetic aperture radar (SAR) can image the Earth's surface and map damage in all weather conditions, day and night. However, current SAR damage mapping methods struggle to separate damage from other changes in the Earth's surface. In this study, we propose a novel approach to damage mapping, combining deep learning with the full time history of SAR observations of an impacted region in order to detect anomalous variations in the Earth's surface properties due to a natural disaster. We quantify Earth surface change using time series of Interferometric SAR coherence, then use a recurrent neural network (RNN) as a probabilistic anomaly detector on these coherence time series. The RNN is first trained on pre-event coherence time series, and then forecasts a probability distribution of the coherence between pre- and post-event SAR images. The difference between the forecast and observed co-event coherence provides a measure of the confidence in the identification of damage. The method allows the user to choose a damage detection threshold that is customized for each location, based on the local behavior of coherence through time before the event. We apply this method to calculate estimates of damage for three earthquakes using multi-year time series of \mbox{Sentinel-1} SAR acquisitions. Our approach shows good agreement with observed damage and quantitative improvement compared to using pre- to co-event coherence loss as a damage proxy. 
\end{abstract}


%

\section{Introduction}\label{s:intro}

\IEEEPARstart{I}{n} the wake of major natural disasters, emergency services need a rapid and accurate assessment of the damage over a wide area in order to quickly direct their response and estimate losses. However, damage to infrastructure and communications networks often makes prompt on-the-ground damage assessment difficult or impossible. Under these circumstances, remote sensing can either complement, or provide a useful alternative to, ground-based assessments \cite{voigt_global_2016}.

Assessments of damage due to a natural disaster can be obtained by comparing satellite observations from before and after the event. One approach is the use of change detection on very high resolution optical data (\( \approx \) 50 $\si{cm}$ $\times$ 50 $\si{cm}$ pixels) \cite{dalla_mura_unsupervised_2008, pesaresi_rapid_2007}. However, the utility of optical images for disaster response can be hampered by the need for timely data, requiring cloud free conditions and sufficient solar illumination \cite{brunner_earthquake_2010}.

Satellite-based synthetic aperture radar (SAR) is an imaging technique that offers advantages over optical data by providing images in all weather conditions, day or night (e.g. \cite{ulaby_microwave_2015,rosen_synthetic_2000}). SAR relies on active imaging using microwave (centimeter-scale) wavelengths emitted by the satellite, with the sensor recording the amplitude and phase of the reflected radar pulse to produce images at meter-scale resolution. The availability of SAR images depends only on the orbital parameters of the satellite.

Damage detection using SAR relies on separating normal changes in the radar backscatter properties of the ground (e.g. due to agricultural activities, vegetation growth, snow, rainfall, and even vehicle motion in a car park) from anomalous changes attributed to disaster-induced damage. The changes can be quantified using the coherence of the radar echo between subsequent acquisitions (\cite{zebker_decorrelation_1992}, and see Eq. \ref{eq:coherence}). One current method of mapping damage uses a pair of SAR acquisitions just before the event and a pair of acquisitions that span the event, allowing for a comparison between the amount of ground surface change without any damage to the amount of ground surface change that occurs during the event \cite{yun_rapid_2015}. This method relies on human judgment in setting an appropriate threshold for classifying areas as damaged, usually assumed to be constant for all locations in the SAR scene, as well as for selecting suitable pre-event acquisitions \cite{yun_rapid_2015}.

There are now satellite SAR missions with revisits on a time-scale of days, and many parts of the Earth have repeat observations by the same satellite constellation going back several years. These developments allow for the possibility of using pre-event multi-year time series to separate out regularly occurring anthropogenic and natural surface changes from changes caused by a given natural disaster. These data have only recently begun to be exploited by researchers for damage mapping purposes \cite{karimzadeh_sequential_2018, washaya_coherence_2018, jung_damage-mapping_2018}. The desire and opportunity to perform damage classification on large and complex SAR data sets motivates us to explore the use of deep learning techniques. 

\IEEEpubidadjcol 

Deep learning relies on feeding input data through multiple layers of non-linear parametrized functions, also known as a deep learning architecture, to transform input data into desired outputs which can be used for regression or classification \cite{3blue1brown_but_2017, lecun_deep_2015, goodfellow_deep_2016}. In supervised deep learning, the function parameters are optimised, during a process known as training, to minimise the misfit between the functions' output and known training data (\textit{ground truth}). For example, in image classification, the function input is an image with a known classification (e.g. ``dog'', ``cat'', ``tree'' etc.), and the function outputs are the probabilities of the image having each classification, with the set of possible classifications finite and fixed. The functions' parameters are optimized to maximize the probabilities assigned to correct classifications for images in a data set, and the final resulting function can then be used to classify previously unseen images. Generally, in supervised learning, the functions are trained using data from the \textit{training set} and then evaluated on a separate data set, the \textit{validation set}, which is unseen during training to ensure the learned functions represent generalizable rules, rather than just a memorization of the training set.

Deep learning has proven to be an effective way to extract insights from large data sets with little or no assumptions about the underlying data, and minimal human intervention \cite{lecun_deep_2015}. The growing body of available satellite data has prompted recent work to combine satellite data with deep learning techniques to study volcanoes \cite{anantrasirichai_deep_2019}, fires \cite{kong_long_2018} and flooding \cite{li_urban_2019} among other examples. 

Recurrent neural networks (RNNs) are a type of deep learning architecture particularly well suited to dealing with sequential (e.g. time-ordered) data \cite{lipton_critical_2015,olah_understanding_2015}. RNNs have been applied to a wide range of tasks, from predicting the next character in a word \cite{sutskever_sequence_2014} to precipitation forecasting \cite{shi_convolutional_2015} and seismic phase association \cite{ross_phaselink_2019}. By training an RNN on a large number of previously observed time series, the network can be used to classify new time series observations and to forecast future time steps. When using an RNN for time series forecasting, the deviation between forecast values and observations can be used for anomaly detection \cite{malhotra_long_2015}.  

The ability of RNNs to learn generalized rules from large time series data sets makes them a good candidate for application to large satellite time series observations, and RNNs have recently begun to be used on satellite data for tasks such as forecasting \cite{shi_convolutional_2015}, classification \cite{ndikumana_deep_2018} and anomaly detection \cite{kong_long_2018}. 

In this study, we frame the damage detection problem as one of detecting anomalies in sequential InSAR coherence time series. We train an RNN on a time series of sequential InSAR coherence data taken before a damage event, then use the trained RNN to make a probabilistic forecast for the co-event InSAR coherence (i.e. the coherence of the radar echo between pre- and post-event acquisitions). The probabilistic nature of the forecast allows us to capture the distribution of the coherence values we expect for each location in the absence of any damage. We then calculate the number of standard deviations of the forecasted distribution between the forecasted mean and the observed co-event coherence value for each point in the region of study. The number of standard deviations between the forecasted mean and observed values is used to quantify how anomalous each coherence value is, with anomalously low coherence values attributed to damage. The use of a probabilistic forecast for each pixel allows us to create a location-dependent threshold for damage which depends on the specific time series characteristics of that location.

In what follows, we summarize the underlying SAR methodology and give a brief overview of previous work on using SAR for damage mapping (Section \ref{s:background}). We then discuss how damage mapping can be formulated in terms of a machine learning problem, and present our method for deploying recurrent neural networks for damage detection (Section \ref{s:approach}). We apply our method to three earthquakes, which had either substantial building damage or surface rupture (Section \ref{s:data}): 
\begin{itemize}
    \item The August 24, 2016 \Mw{6.2} central Italy earthquake
    \item The November 12, 2017 \Mw{7.3} Iran-Iraq earthquake
    \item The July 2019 \Mw{6.4} and \Mw{7.1} Ridgecrest, California, USA earthquakes.
\end{itemize}

Through these examples, we illustrate how combining a long pre-event SAR time series with RNN-based anomaly detection can improve results compared to an existing SAR damage mapping method (Section \ref{s:results}). We discuss the strengths and limitations of our proposed method (Section \ref{s:discussion}) then present conclusions and outline potential further work (Section \ref{s:conclusion}). Further details of our deep learning architecture as well as the satellite data, damage assessments, and example code used in this study are presented in the supplementary materials. 

\section{Background and Previous Work}\label{s:background}

\subsection{Synthetic Aperture Radar}\label{s:background:sar}
Synthetic Aperture Radar (SAR) is a coherent active imaging method operating at microwave wavelengths used for mapping the Earth's surface \cite{ulaby_microwave_2015}. The method relies on satellite-based illumination of the ground with 1--30 cm wavelength microwaves, then recording the amplitude and phase of the reflected wave. In our work, we begin with processed full-resolution data known as \textit{single look complex} (SLC) images. Each SLC pixel in the image corresponds to a region on the Earth's surface and records the amplitude and phase of the radar echo from that region. 
The reflected wave depends on the properties of the Earth's surface, with the echo being a combination of the coherent sum of the backscatter from all of the reflectors within an SLC pixel, or \textit{resolution element} (e.g. Section 3.12.2 of \cite{simons_interferometric_2015}), as well as delays accrued during propagation through the atmosphere (e.g. Section 3.12.4.2 of \cite{simons_interferometric_2015}).

\subsection{Change Detection using Synthetic Aperture Radar}\label{s:background:changedetection}

Changes in the imaging or viewing geometry, surface roughness, and dielectric properties of the ground within a resolution element will affect the measured radar return \cite{zebker_decorrelation_1992, jordan_surface_2020}.
For example, the collapse of buildings changes the path length travelled by the radar wave and randomly rearranges the radar reflectors within a given SLC pixel, leading to a random change in each SLC pixel's phase.

Comparing SAR images of the same point on Earth from the same satellite taken at different times provides proxies for changes in the Earth's surface. These measurements can be classified as coherent or incoherent, depending on whether or not the SAR phase is used \cite{jung_evaluation_2020}. In this study, we focus on coherent change detection where the change between two SAR acquisitions can be quantified by calculating the magnitude of the complex correlation coefficient, also known as the \textit{interferometric coherence} or simply \textit{coherence}, between the two complex SAR signals. For a given SLC pixel, coherence is defined as:
\begin{equation}
    \label{eq:coherence}
    \gamma_{i,j} = \left\| \frac{| \langle \Gamma_{i}\Gamma_{j}^* \rangle|}{\sqrt{ \langle|\Gamma_{i}|^2\rangle \langle|\Gamma_{j}|^2\rangle}} \right\|,
\end{equation}
where $\Gamma_{i}$ is the complex amplitude and phase for SAR acquisition at time step $i$, $^*$ represents complex conjugation, and $\langle \rangle$ denotes an ensemble average, generally approximated as a local spatial average (e.g. see Section 3.12.2.5 of \cite{simons_interferometric_2015}). $\gamma_{i,j}$ is known as the coherence of the signal between SAR acquisitions at time steps $i$ and $j$. This measure incorporates information about changes in both the amplitude and phase of the SAR signal. The coherent nature of SAR means that it is possible to sense changes on the scale of the radar wavelength (1-30 cm) when using phase information, allowing for very sensitive change detection compared to most optical data.

The use of a local spatial average in the coherence calculation means that the resolution of the coherence image is necessarily lower than the original SAR SLC image, as multiple pixels in the SLC image (SLC pixels) are used to calculate a single pixel in the coherence image (coherence pixel). Unless stated otherwise, the term \textit{pixel} refers to coherence pixels for the rest of this paper.

For completely coherent echos $\gamma_{i,j}=1$, whereas $\gamma_{i,j} = 0$ implies that the two echos are completely uncorrelated (a low or zero coherence value is also known as decorrelation). Stable, concrete structures, for example, will reflect radiation in the same manner through time, and thus exhibit high coherence, whereas bodies of water, which change their radar scattering properties on a time-scale of less than a second, will completely decorrelate \cite{bamler_synthetic_1998}.

Coherence for a given pixel will tend to decrease with the time between SAR acquisitions due to natural changes in the Earth's properties, with the rate of decrease depending on the rate of change of the Earth's backscattering properties at length scales similar to the radar wavelength \cite{zebker_decorrelation_1992}. However, the presence of seasonal effects such as snow can also lead to seasonal coherence variations as the ground surface is covered and uncovered, and rainfall can lead to sudden drops in coherence \cite{jordan_surface_2020}. The time between the two acquisitions, known as the temporal baseline, is therefore an important indicator of how much coherence loss to expect. Increasing the spatial separation, known as the spatial baseline, between the SAR sensor's image acquisition position for repeat images will also lead to a decrease in coherence \cite{zebker_decorrelation_1992}. Currently orbiting satellites have tight orbital control, such that spatial baseline decorrelation is a less significant problem than it was for previous generations of sensors. 

A spatial image of coherences calculated from two SAR acquisitions, acquired at different times, allows for mapping of changes in the Earth's surface properties, on the scale of the radar wavelength. For example, Simons \textit{et al.} \cite{simons_coseismic_2002} used the spatial pattern of low coherence to map the location of fault surface rupture due to the 1999 \Mw{7.1} Hector Mine, California earthquake. However, decorrelation effects from regularly occurring natural processes often occur together with those induced by damage events \cite{jung_damage-mapping_2018}. 
Within a coherence image spanning an earthquake (co-event coherence), we may detect decorrelation due to collapsed buildings as well as, for example, agricultural activity, vegetation growth and the changing position of vehicles in a car park, making the isolation of damage effects challenging. 

The need to separate changes in surface properties due to damage from other changes motivates the framing of this problem as one of anomaly detection. If we are able to identify the nominal distribution of coherence (at a given temporal baseline) for each pixel before any damage has occurred, we can then identify which pixels have an anomalously low co-event coherence with respect to their pre-event distribution and use the presence of anomalous coherence as a proxy for damage. This nominal distribution may be a complicated function of underlying physical properties, and may not be stationary in time.  

\begin{figure*}
\includegraphics[width=\textwidth]{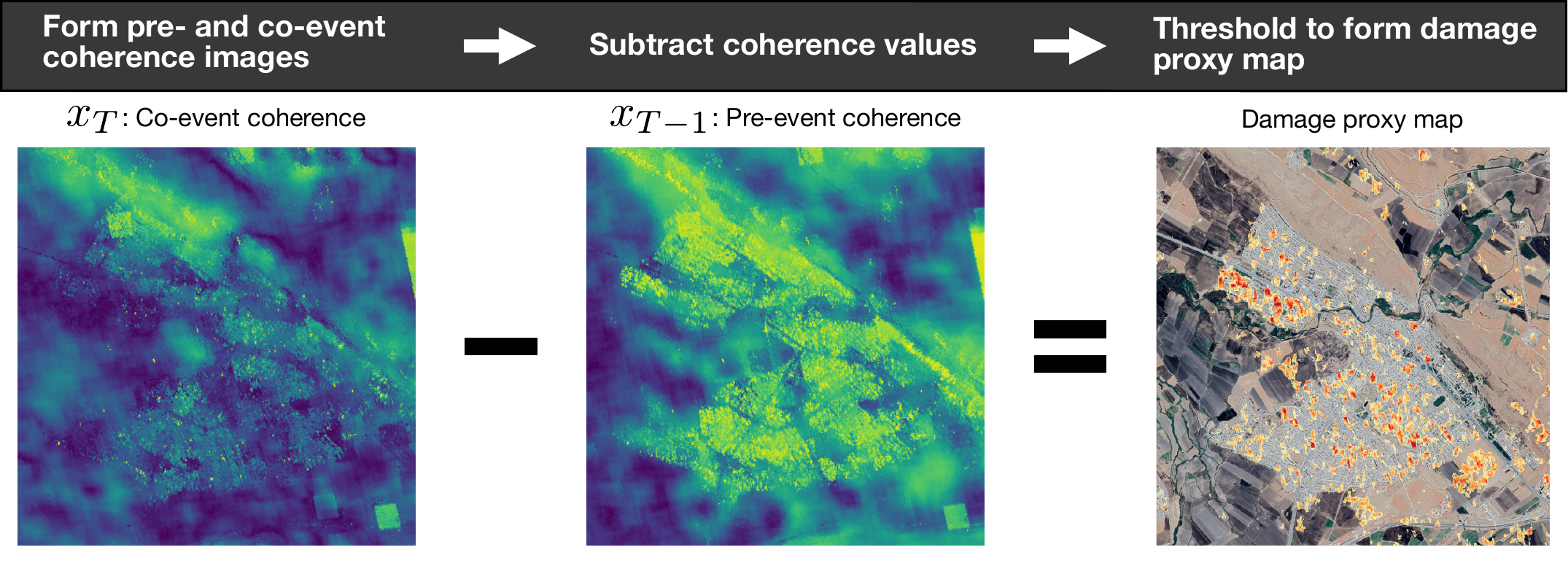}
\caption{\label{fig:ccd_method_summary} Schematic of the existing Coherence Change Detection (CCD) method for damage mapping \cite{yun_rapid_2015}, presented for the town of Sarpol-e-Zahab, damaged during the November 2017 Iran-Iraq earthquake. A pre-event coherence image ($x_{T-1}$) is subtracted from the co-event coherence image ($x_T$) in order to calculate the coherence loss. The coherence loss is thresholded and plotted to produce a damage proxy map. Optical data from Google, CNES/Airbus, taken July 27th 2020.}
\end{figure*}

One way to characterize the pre-event coherence is to calculate the coherence between two SAR SLCs acquired as close as possible before the event. The co-event coherence can then be compared to the pre-event coherence and the magnitude of the relative coherence loss can be used to identify areas where the coherence has dropped anomalously. Generally, a threshold for the amount of coherence loss required for a pixel to be marked as damaged is chosen manually, by including areas where it is known that no damage occurred and setting the threshold so that these undamaged areas are correctly classified. This method is sometimes known as Coherence Change Detection (CCD) (e.g. see \cite{yun_rapid_2015, fielding_surface_2005, bouaraba_robust_2012, geudtner_flood_1996} and Fig. \ref{fig:ccd_method_summary}), and is based on the assumption that the calculated pre-event coherence image is a good representation of the normal pre-event coherence. In cases where coherence between sequential SAR acquisitions has a high variance (i.e. there is a lot of variation in the amount of surface change for a given temporal baseline), a single pre-event coherence image will not be a good characterization of the pre-event coherence distribution for the given temporal baseline, and the CCD damage map is likely to be noisy.

To better characterize the pre-event coherence, researchers have begun using the long time series of regular SAR acquisitions that are increasingly available \cite{karimzadeh_sequential_2018, jung_damage-mapping_2018}. By calculating the coherence between sequential SAR acquisitions, the mean and standard deviation of the sequential pre-event coherence can be calculated for each pixel. The number of standard deviations between the mean pre-event coherence and the co-event coherence can then be used to detect anomalous co-event decreases in coherence \cite{washaya_coherence_2018,olen_mapping_2018}. These methods rely on characterising the pre-event coherence with a single distribution through time for each pixel, which can cause problems when the coherence distribution varies substantially through time, for example due to changing precipitation with the seasons.

Additional information can be gained by calculating the coherence between all possible SAR pairs, leading to coherence images with a wide range of temporal baselines \cite{monti-guarnieri_coherent_2018, jordan_surface_2020}. These coherence values can be used to estimate the parameters, for each pixel, of a model for the various contributors to temporal decorrelation \cite{jung_coherent_2016, jung_damage-mapping_2018}. This model can then be used to detect anomalies in coherence images which span the event. Similar to the mean and standard deviation method, these methods generally rely on inferring a single set of physical parameters for each pixel, without taking into account the possible variation of these parameters through time.

Supervised machine learning has also been used for damage mapping with SAR data, using comprehensive damage assessments, often available several months after major events, as ground truth to train damage classifiers \cite{wieland_learning_2016, endo_new_2018, li_urban_2019}. While supervised machine learning approaches avoid the problem of manually selecting a uniform damage threshold in CCD, they rely on extensive ground truth damage assessment data for training. Additionally, if the damage classifiers are to be useful for future events, the trained classifiers must be applied to new areas and it is unclear to what extent this training readily transfers to totally different regions of the Earth's surface. 

In our work, we seek to make use of all available SAR data before an event in order to make a deep learning-based, time-dependent forecast of a co-event coherence distribution that we would expect without any damage event. This approach allows us to detect anomalous changes in coherence. As we only use ground truth damage data to quantify our damage detection algorithm, and not for training, our method does not depend on ground truth damage data.

\section{Proposed Approach}\label{s:approach}

\begin{figure*}
\includegraphics[width=\textwidth]{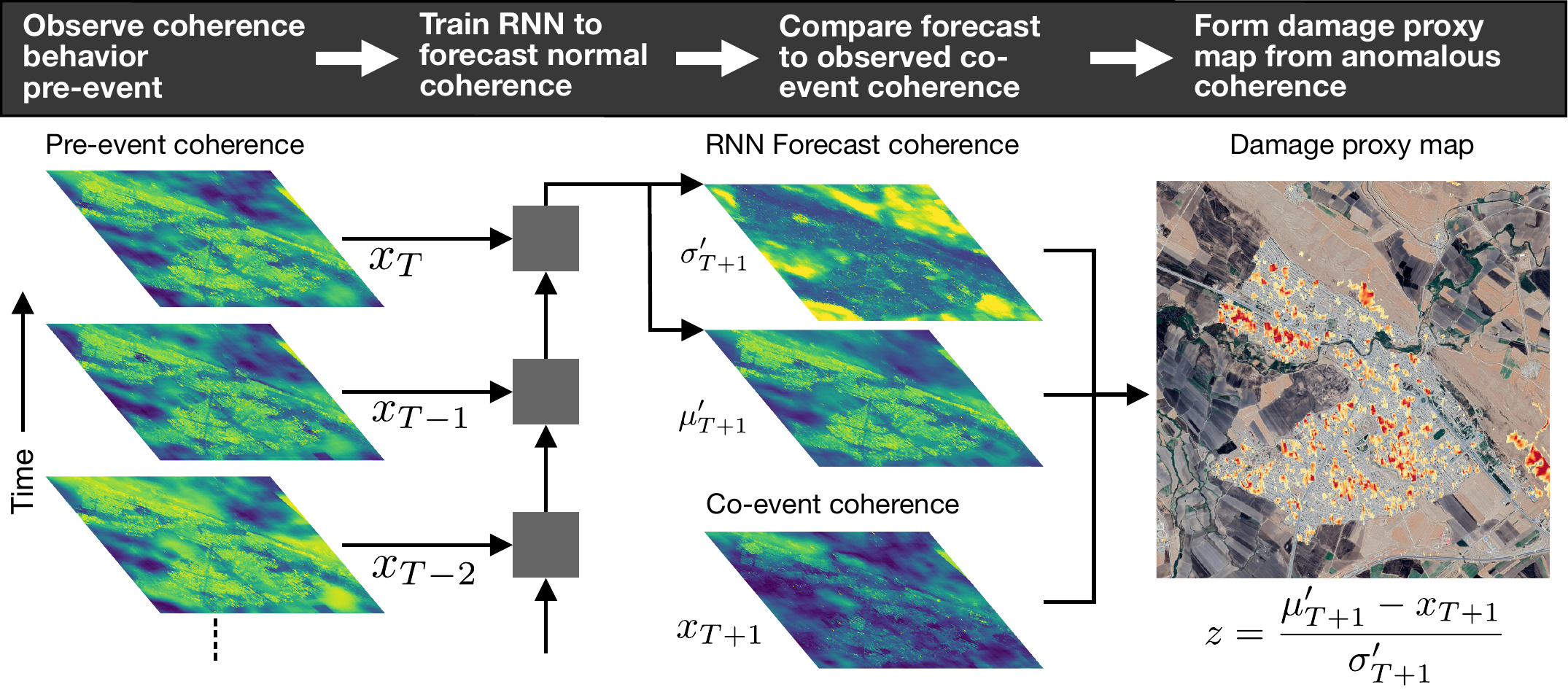}
\caption{\label{fig:method_summary} Schematic of our proposed recurrent neural network (RNN) method presented for the town of Sarpol-e-Zahab, damaged during the November 2017 Iran-Iraq earthquake. The transformed coherence values ($x$) are used to train a recurrent neural network to make a Gaussian forecast of the co-event coherence with mean $\mu'_{T+1}$ and standard deviation $\sigma'_{T+1}$. The forecast is compared with the observed co-event coherence, $x_{T+1}$, to calculate the z-score, $z$ (see Eq. \ref{eq:z_score}). The z-score is thresholded and plotted to produce a damage proxy map. A more detailed illustration of the neural network architecture can be found in Fig. \ref{ext-fig:graphical_model}. Optical data from Google, CNES/Airbus, taken July 27th 2020.}
\end{figure*}

\subsection{Notation}

We have a total of $T+2$ SAR acquisitions, ordered in time and indexed from 0 to $T+1$; the last acquisition, $T+1$, is post-event, while all others are pre-event. Between all pairs of consecutive acquisitions we compute the coherence values which we map to an unbounded space using a logit transform on the squared coherence (discussed below, see Eq. \ref{eq:inverse-sigmoid}). We write the coherence between time steps $t-1$ and $t$ as $\gamma_{t,t-1}$ and the transformed coherence as $x_t$. Throughout the rest of the paper, \textit{coherence} refers to the transformed coherence unless otherwise stated. Let $x_{\leq T} = \{ x_{1}, ..., x_{T}\}$ denote the sequence of $T$ pre-event sequential coherence values for a given coherence pixel location, with $x_{T+1}$ the co-event coherence. Additionally, let $\mathcal{D}_t$ denote the collection of $M$ coherence sequences that we have available for training (i.e. coherence sequences from $M$ different coherence pixels), each containing $T$ pre-event coherence values. We also have $\mathcal{D}_f$, the set of coherence sequences on which we wish to perform forecasting and anomaly detection for $x_{T+1}$, which contains $T$ pre-event ($x_{\leq T}$) and one co-event ($x_{T+1}$) coherence values for each pixel.  

Our goal is to train a model that can capture the range of possible behaviors across time for the sequences in $\mathcal{D}_t$. We can then use the model to make a forecast, $x'_{T+1}$ (we use $'$ to denote a forecasted value), for each co-event coherence pixel in $\mathcal{D}_f$, based on the coherence time series of that pixel. We can detect anomalies by comparing the forecast with the ground truth co-event coherence value, $x_{T+1}$, at each pixel, mapping anomalous changes in ground surface properties that have occurred between SAR acquisitions $T$ and $T+1$. Note that the model does not see $x_{T+1}$, or any damage data, during training. 

\subsection{Recurrent Neural Networks}

Motivated by the sequential nature of our data, we use a recurrent neural network (RNN) as our model for forecasting the coherence time series. RNNs are a class of models frequently used on sequential data for machine learning tasks such as speech recognition, machine translation, motion tracking and time series classification and forecasting \cite{lipton_critical_2015,olah_understanding_2015}. RNNs maintain a fixed-length hidden state vector, $h_{t}$, that summarizes a sequence up to time $t$, and is updated at every time step with observations:
\begin{equation}
    h_{t} = f_{\phi}(h_{t-1}, x_{t}),
\label{eq:hidden_state}
\end{equation}
where $f$ is a deterministic function, learned during training and parametrized by $\phi$, and $x_t$ is the transformed coherence (Eq. \ref{eq:inverse-sigmoid}), calculated from SAR data, at time step $t$ for a given pixel. Forecasting future values involves another function $g$ parametrized by $\psi$:
\begin{equation}
\label{eq:fcnn}
    x'_{t+1} = g_{\psi}(h_{t}).
\end{equation}

In general one can optimize for parameters $\phi$ and $\psi$ to minimize some loss, or cost function, between the model forecast and the coherence ground truth (here the transformed coherence values, see Eq. \ref{eq:inverse-sigmoid}), for example the mean-squared error: 
\begin{equation}
\label{eq:mse}
    \phi^*, \psi^* = \arg \min_{\phi, \psi} \sum_{x_{\leq T} \in \mathcal{D}_t} \sum_{t=1}^{T} (x_{t} - x'_{t})^2.
\end{equation}

RNNs use neural networks as function approximators for $f$ and $g$, and this optimization can be solved with some form of gradient descent (e.g. \cite{kingma_adam_2014}). See Section \ref{ext-a:rnn_details} for more details. 

\subsection{Probabilistic Formulation}

We aim to forecast the probability distribution over all possible values given the pre-event coherence values. This probabilistic forecast lets us evaluate the probabilities that our model assigns to the coherence values that are actually observed. We locate anomalies by identifying coherence values that have a low probability of occurring given the previous observations. To make a probabilistic forecast, we modify $g$ in Eq. \ref{eq:fcnn} to output the parameters of a probability distribution instead of a single value. In this work we use a Gaussian output probability, so we have:
\begin{equation}
\label{eq:gauss_fcnn}
    [\mu'_{t+1}, \sigma'_{t+1}] = g_{\psi}(h_{t}),
\end{equation}
where $\mu'_{t+1}$ and $\sigma'_{t+1}$ are the forecast mean and standard deviation respectively.
The probability our model assigns to the ground truth co-event coherence, $x_{t+1}$, is then:
\begin{equation}
\label{eq:gauss}
\begin{aligned}
    p(x_{t+1} ; \mu'_{t+1}, \sigma'_{t+1}) =& (2 \pi \sigma_{t+1}^{'2})^{-\frac{1}{2}} \\ & \times\exp\left(-\frac{\left(x_{t+1}-\mu'_{t+1}\right)^2}{2\sigma_{t+1}^{'2}}\right).
\end{aligned}
\end{equation}
Instead of minimizing the mean-squared error, as shown in Eq. \ref{eq:mse}, the probabilistic forecast allows us to maximise the probability that our model assigns to ground truth sequences in $\mathcal{D}_t$. Computationally, probability maximization is best achieved by minimizing the negative log-likelihood that the model assigns to ground truth sequences in $\mathcal{D}_t$:
\begin{equation}
\label{eq:loglikelihood}
\begin{aligned}
    \phi^*, \psi^* & = \arg \min_{\phi, \psi} \sum_{x_{\leq T} \in \mathcal{D}_t} -\log_e p(x_{\leq T}) \\
    & = \arg \min_{\phi, \psi} \sum_{x_{\leq T} \in \mathcal{D}_t} \sum_{t=1}^T -\log_e p(x_t | x_{<t}),
\end{aligned}
\end{equation}
where in the second step we factorize the conditional probabilities using the relationship $p(x_{\leq T}) = \Pi_{t=1}^T p(x_t | x_{<t})$. In our case, all of the information from previous elements in the time series is summarised by the $\mu'$ and $\sigma'$ terms given by the forecast in Eq. \ref{eq:gauss_fcnn}, so we have that $p(x_t|x_{<t}) = p(x_t;\mu'_t,\sigma'_t)$. Therefore, combining Eq. \ref{eq:gauss} and Eq. \ref{eq:loglikelihood} we have:
\begin{equation}
\label{eq:gauss_nll}
\begin{aligned}
    \phi^*, \psi^* & = \arg \min_{\phi, \psi} \\& \sum_{x_{\leq T} \in \mathcal{D}_t} \sum_{t=1}^T \left( \frac{1}{2}\log_e(2\pi \sigma'^2_{t}) + \frac{(x_t - \mu'_t)^2}{2\sigma'^2_{t}} \right).
\end{aligned}
\end{equation}

Eq. \ref{eq:gauss_nll} gives us a loss function that takes into account both the mean and standard deviation of the forecast, allowing us to optimize a probabilistic forecast for the coherence. We can optimize for the parameters in Eq. \ref{eq:gauss_nll} using some form of gradient descent (e.g. \cite{kingma_adam_2014}). .

Note that a Gaussian distribution assigns nonzero probability everywhere in $\mathbb{R}$, a distribution that is inconsistent with coherence as defined in Eq. \ref{eq:coherence}, which is by definition bounded (i.e. $\gamma_{t-1,t} \in [0, 1]$). We therefore transform the coherence to an unbounded space before training the RNN. We choose the inverse-sigmoid transform (also known as the logit transform) on the square of the coherence values:  
\begin{equation}
\label{eq:inverse-sigmoid}
    x_t = S^{-1}(\gamma^2_{t-1,t}) = \log_e\left(\frac{\gamma^2_{t-1,t}}{1 - \gamma^2_{t-1,t}}\right),
\end{equation}
which maps the domain from $[0, 1]$ to $(-\infty,\infty)$. 
This choice of transform is motivated by near mathematical equivalence between the logit transform of coherence squared and the logarithm of the variance of the interferometric phase, see Section \ref{ext-a:phase_var} for more details. We refer to the new unbounded space as the logit space. Our model will then forecast Gaussian distributions over the unbounded logit space.

\subsection{Model and Training Details}

The RNN model we use in this work (represented by $f_\phi$ in Eq. \ref{eq:hidden_state}) is called a gated recurrent unit (GRU), chosen for its ability to learn long-term dependencies in time series \cite{cho_properties_2014}. The hidden state output from $f_\phi$ is fed into a feed-forward neural network, represented by $g_\psi$ (Eq. \ref{eq:gauss_fcnn}) which then outputs the parameters of the forecast distribution. To find the optimum model parameters (Eq. \ref{eq:gauss_nll}), we train the model using the Adam optimizer \cite{kingma_adam_2014}. See Section \ref{ext-a:rnn_details} for more details of the model and training, as well as further references. An implementation of our deep learning model can be found on GitHub: \url{https://github.com/olliestephenson/dpm-rnn-public}.

\subsection{Anomaly Detection for Co-event Coherence}

To construct a proxy for damage we normalize the difference between the forecast co-event mean, $\mu'_{T+1}$, and the observed co-event coherence, $x_{T+1}$, by the standard deviation of the forecast $\sigma'_{T+1}$. This quantity is termed the z-score, which we define as:

\begin{equation}\label{eq:z_score}
    z = \frac{\mu'_{T+1} - x_{T+1}}{\sigma'_{T+1}}.
\end{equation}

Note that we have switched the order of $\mu'_{T+1}$ and $x_{T+1}$ terms compared to usual definition of the z-score. With this definition, a large positive z-score implies that the coherence is many standard deviations below the forecasted coherence, i.e. we have an anomalous drop in coherence, possibly due to damage. We use this definition of the z-score as the basis of our proxy for damage.  

\section{Data}\label{s:data}

\subsection{Coherence Calculation}\label{s:data:coherence}

We use data from the Copernicus \mbox{Sentinel-1} satellites, a pair of C-band SAR satellites operated by the European Space Agency. We download freely available Level-1 Single Look Complex (SLC) images acquired in interferometric wideswath mode \cite{european_space_agency_single_nodate}. We then create a coregistered stack of SLCs covering the region of interest. To generate coherence values, as defined in Eq. \ref{eq:coherence}, we average over a rectangle, or chip, of SLC pixels. In this case we use a chip of 15 SLC pixels in range (across the satellite track) and 5 in azimuth (along the satellite track) corresponding to a region of approximately $50~\si{m}$ by $70~\si{m}$.  Note that the resolution in range (across-track) of \mbox{Sentinel-1} SLCs is higher than in the azimuth (along-track) direction. As stated above, the use of a chip to calculate coherence means that the coherence map is lower resolution than the SLC image as each coherence pixel contains information from a $50~\si{m}$ by $70~\si{m}$ area.

For each study area, we produce two separate coherence data sets: one for training the network ($\mathcal{D}_t$), and one for forecasting purposes ($\mathcal{D}_f$). We construct the training data set to have a large number of pixels drawn from a wide area surrounding the area of interest, while the forecasting data set focuses just on the area of interest to be mapped. More details on how these data sets are constructed can be found in Section \ref{ext-a:study_areas}.

\subsection{Study Areas}\label{s:data:areas}
In this study, we consider three earthquakes:
\subsubsection{August 24, 2016 \Mw{6.2} central Italy earthquake}
This event destroyed much of the town of Amatrice in central Italy. \cite{united_states_geological_survey_m_2016}. The Copernicus Emergency Management Service produced a damage map assessing the damage level of every building in the town \cite{copernicus_emergency_management_service_emsr177_2016}. This comprehensive damage assessment allows us to quantitatively validate the RNN and CCD methods against the known damage levels.
\subsubsection{November 12, 2017 \Mw{7.3} Iran-Iraq earthquake}
This event damaged the city of Sarpol-e-Zahab on the Iran-Iraq border \cite{united_states_geological_survey_m_2017}. The United Nations Institute for Training and Research (UNITAR) produced a damage map for Sarpol-e-Zahab in the wake of the earthquake \cite{united_nations_institute_for_training_and_research_damaged_2017}, allowing for a qualitative test of our damage proxy map and comparison with the CCD method. 
\subsubsection{July 2019 \Mw{6.4} and \Mw{7.1} Ridgecrest, California, USA earthquakes}
To explore the ability of our method to capture other forms of anomalous ground disturbances, we also consider the Ridgecrest earthquakes which struck the Mojave desert, California, in early July 2019. The Ridgecrest sequence contained two earthquakes with substantial surface rupture tens of kilometers long; a \Mw{6.4} event on July 4th, and, 34 hours later, a \Mw{7.1} event \cite{ross_hierarchical_2019,kendrick_geologic_2019, ponti_documentation_2020}. The earthquakes also caused liquefaction, small rock falls and minor damage to buildings \cite{stewart_preliminary_2019, zimmaro_liquefaction_2020, hough_nearfield_2020}. The mapping of surface ruptures and location of liquefaction allows us to qualitatively compare the damage map to the location of known ground surface changes. 

In Section \ref{ext-a:study_areas}, we give more detailed information about these three earthquakes and the available data for each event.

\section{Results}\label{s:results}
We present damage proxy maps for the coherence change detection (CCD) and our proposed RNN methods, then use available independent damage data to validate the efficacy of each method. For each method, we calculate a numerical damage proxy for every pixel, then threshold that damage proxy to create damage proxy maps. For Sarpol-e-Zahab and Ridgecrest, the limited nature of the ground truth data only allows for a qualitative comparison between the methods. For Amatrice, however, more comprehensive ground truth allows us to carry out a quantitative comparison. For Sarpol-e-Zahab, we also explore the forecasts the RNN makes through time for pixels in different locations.
We find that the RNN method yields qualitative and quantitative improvements over the the CCD method.

\subsection*{August 24, 2016 \Mw{6.2} earthquake, Amatrice, Italy}

\begin{figure*}
    \centering
    \includegraphics[width=\textwidth]{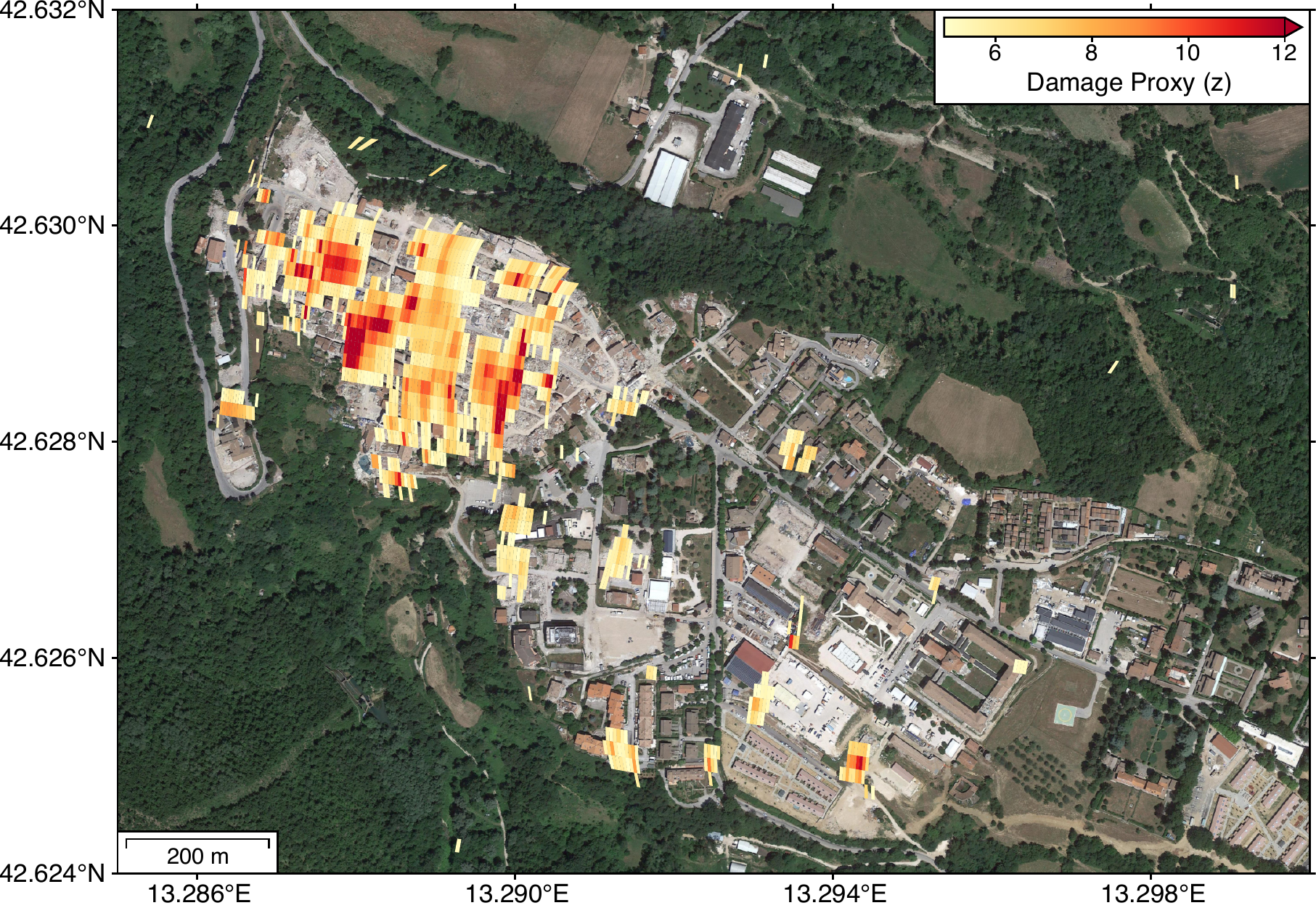}
    \caption{RNN DPM for the town of Amatrice, Italy, badly damaged during the 2016 \Mw{6.2} central Italy earthquake. The center of the town, which was largely destroyed, is clearly highlighted by elevated damage proxy values towards the top left of the map. Z-score values below 4.93 (chosen from the F$_{0.5}$ score, Eq. \ref{eq:f_beta}) are masked, values above 12 are set to red as indicated by the color bar. Ground truth damage data is presented in Figure \ref{ext-fig:amatrice_ground}. Optical imagery from Google, taken July 6th 2017.}
    \label{fig:amatrice_rnn_dpm}
\end{figure*}

Fig. \ref{fig:amatrice_rnn_dpm} shows the RNN method applied to mapping the damage in the town of Amatrice due to the August 24, 2016 central Italy earthquake. We use ground truth damage data from the Copernicus Emergency Management Service \cite{copernicus_emergency_management_service_emsr177_2016} to choose an optimum threshold for damage (discussed below), and mask values below that threshold. Details of the damage data are presented in Section \ref{ext-a:study_areas}. The ground truth damage data also allows for a direct quantitative comparison between the CCD and RNN methods.

\begin{figure*}
\includegraphics[width=\textwidth]{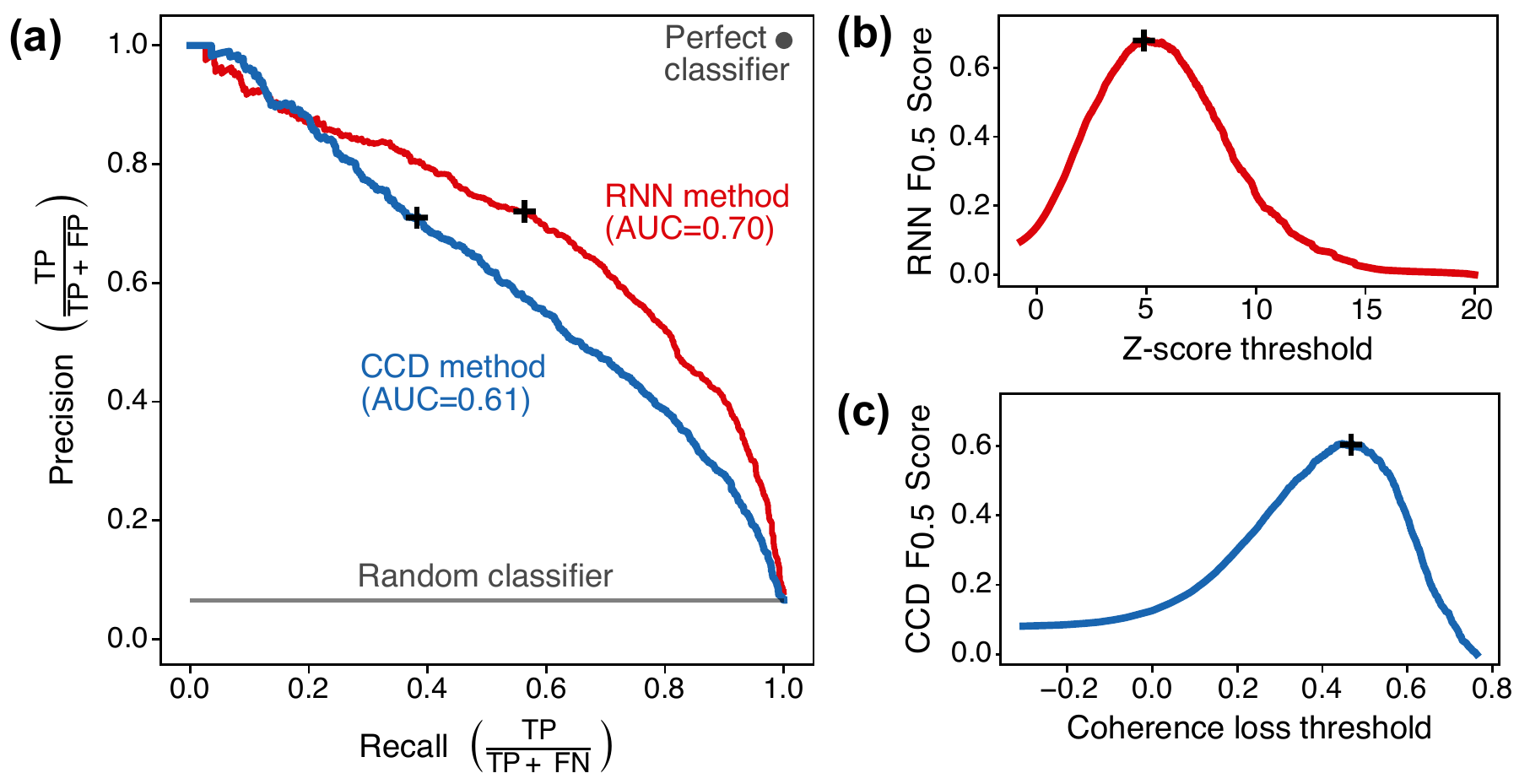}
\caption{\textbf{(a)} Precision-recall curves for Amatrice damage proxy maps using the CCD (blue line) and RNN (red line) methods. For a perfect classifier we can choose a threshold that gives precision and recall equal to one, indicated in the top right corner. A random classifier gives a constant precision, equal to the fraction of the data set that is truly damaged, with recall varying with the threshold, indicated by the grey horizontal line at the bottom of the plot. The larger area under curve (AUC) for the RNN method indicates improved performance. The black crosses show the position of maximum $\text{F}_{0.5}$ score identified in figures \textbf{(b)} and \textbf{(c)}. \textbf{(b)} $\text{F}_{0.5}$ score (see Eq. \ref{eq:f_beta}) for varying z-score damage thresholds using the RNN method. \textbf{(c)} $\text{F}_{0.5}$ score for varying coherence loss damage thresholds using the CCD method.}
\label{fig:amatrice_pr_f05}
\end{figure*}

We seek to classify each coherence chip as either ``damaged'' or ``undamaged'' and compare the classification ability of the CCD and RNN methods. Comparing the classifiers relies on assigning each chip a score (the z-score for the RNN method and the coherence loss for the CCD method), setting a threshold for damage, and then comparing the  damaged/undamaged classifications with the ground truth. For every set of classifications, we have four categories: assigned damaged and truly damaged (true positive), assigned damaged but actually undamaged (false positive), assigned undamaged and truly undamaged (true negative) and assigned undamaged but actually damaged (false negative). 

In this case the ground truth damage data is building footprints, each with a damage score, which we separate into ``damaged'' and ``undamaged'' classes (see Section \ref{ext-a:study_areas}). To determine the ground truth associated with each coherence chip, we calculate the proportion of each chip's area that is occupied by damaged buildings. Coherence chips that have at least one third of their area (roughly 1200 $\text{m}^2$, see the discussion in Section \ref{s:discussion}) occupied by the footprints of damaged buildings, we assign to be truly damaged. Note that as the radar is side-looking, the 3D nature of the buildings means their radar footprint does not exactly match their ground footprint, a fact that we do not take into account in this work. 

To compare the two methods quantitatively, we use a standard precision-recall curve \cite{davis_relationship_2006} (Fig. \ref{fig:amatrice_pr_f05}). We calculate precision (the fraction of chips classified as damaged that are actually damaged) and recall (the fraction of truly damaged chips that are classified as damaged) for a range of damage proxy thresholds for each method. For a general classifier that is imperfect but better than random, precision and recall will trade off against one another. For example, with a high threshold, only a few points will be classified as damaged, and most of these will be truly damaged, leading to a high precision. However, with a high threshold the recall is low as most truly damaged points are incorrectly classified as undamaged. A low threshold means that many of the damaged points are above the threshold, however there are also many false positives, leading to low precision and high recall. As our classes are unbalanced (there are many more undamaged points than damaged points) the precision-recall curve is preferred over the receiver operating characteristic (ROC) curve that is also used to assess the quality of classifiers \cite{davis_relationship_2006}.

Different classification methods can be quantitatively compared by calculating the area under the precision-recall curve (known as PR AUC). A perfect classifier will have an area of unity, with better algorithms having PR AUCs closer to this value. The PR AUC for a random classifier will be equal to the fraction of the data set that is truly damaged. Note that the PR AUC is distinct from the ROC AUC which is also used to compare classifiers \cite{berrar_caveats_2012}. 
Our PR AUC results presented in Fig. \ref{fig:amatrice_pr_f05}\textbf{(a)} show a clear quantitative improvement when using the RNN method over the CCD method, with a PR AUC of 0.70 for the RNN method and 0.61 for the CCD method. We achieve this improvement using the RNN method in spite of the relatively poor quality training data (see discussion in Section \ref{s:discussion}).

To compute the optimum threshold for damage for each method, we can use the $\text{F}_{\beta}$ score, which is the weighted harmonic mean of the precision and the recall, computed as: 
\begin{equation}\label{eq:f_beta}
    \text{F}_\beta = (1+\beta^2)\frac{\text{precision}\cdot\text{recall}}{(\beta^2\cdot\text{precision}) + \text{recall}}.
\end{equation}
$\text{F}_{\beta}$ will vary with the threshold, and we can choose a threshold that maximizes the score. This weighting considers recall $\beta$ times as important as precision. In our case we set $\beta=0.5$ and thus compute the $\text{F}_{0.5}$ score for all possible thresholds for both methods. Our choice of $\beta$ weights precision as twice as important as recall, based on the assumption that we wish to direct finite emergency response resources to the places most likely to be damaged and thus favour higher precision at the expense of lower recall. At the maximum $\text{F}_{0.5}$ values, both methods have a precision just over 0.7, meaning over 70\% of the points classified as damaged are truly damaged, however the RNN method has recall of 0.56, compared to 0.38 for the CCD method, a clear quantitative improvement. The $\text{F}_{0.5}$ scores are presented in Fig. \ref{fig:amatrice_pr_f05}\textbf{(b)} and \textbf{(c)}, and the optimum threshold and corresponding precision and recall for each method are given in Table \ref{table:opt_thresh}.

Note that we perform the quantification on a pixel-by-pixel basis (using coherence pixels) rather than a building-by-building basis. Building areas can vary greatly, and, for the same level of damage, a small building and a large building can have very different effects on the coherence. Therefore a building-by-building quantification would combine metrics with very different sensitivities, whereas in theory each coherence chip should respond in a more similar way when a given fraction of its area is occupied by buildings with the same level of damage. We also note that in actual deployment scenarios, building footprints may not be available, and we may also be interested in investigating other forms of anomalous surface change (for example fault ruptures and landslides).

\begin{table}
\caption{Optimum threshold and corresponding precision and recall values for both methods using the Amatrice data set, selected using the maximum value of the $F_{0.5}$ score, along with the area under the precision-recall curves (PR AUC)}
 \begin{tabularx}{\columnwidth}{|X|X|X|X|X|}
 
\hline
 Method & Optimum threshold & Optimum precision & Optimum recall & PR AUC \\
 \hline
  RNN & 4.93 & 0.72 & 0.56 & 0.70 \\ 
 \hline
 CCD & 0.47 & 0.71 & 0.38 & 0.61 \\
 \hline
\end{tabularx}
 \label{table:opt_thresh}
\end{table}

\subsection*{November 12, 2017 \Mw{7.3} earthquake, Sarpol-e-Zahab, Iran-Iraq Border}

\begin{figure*}[!t]
\centering
\includegraphics[width=\textwidth]{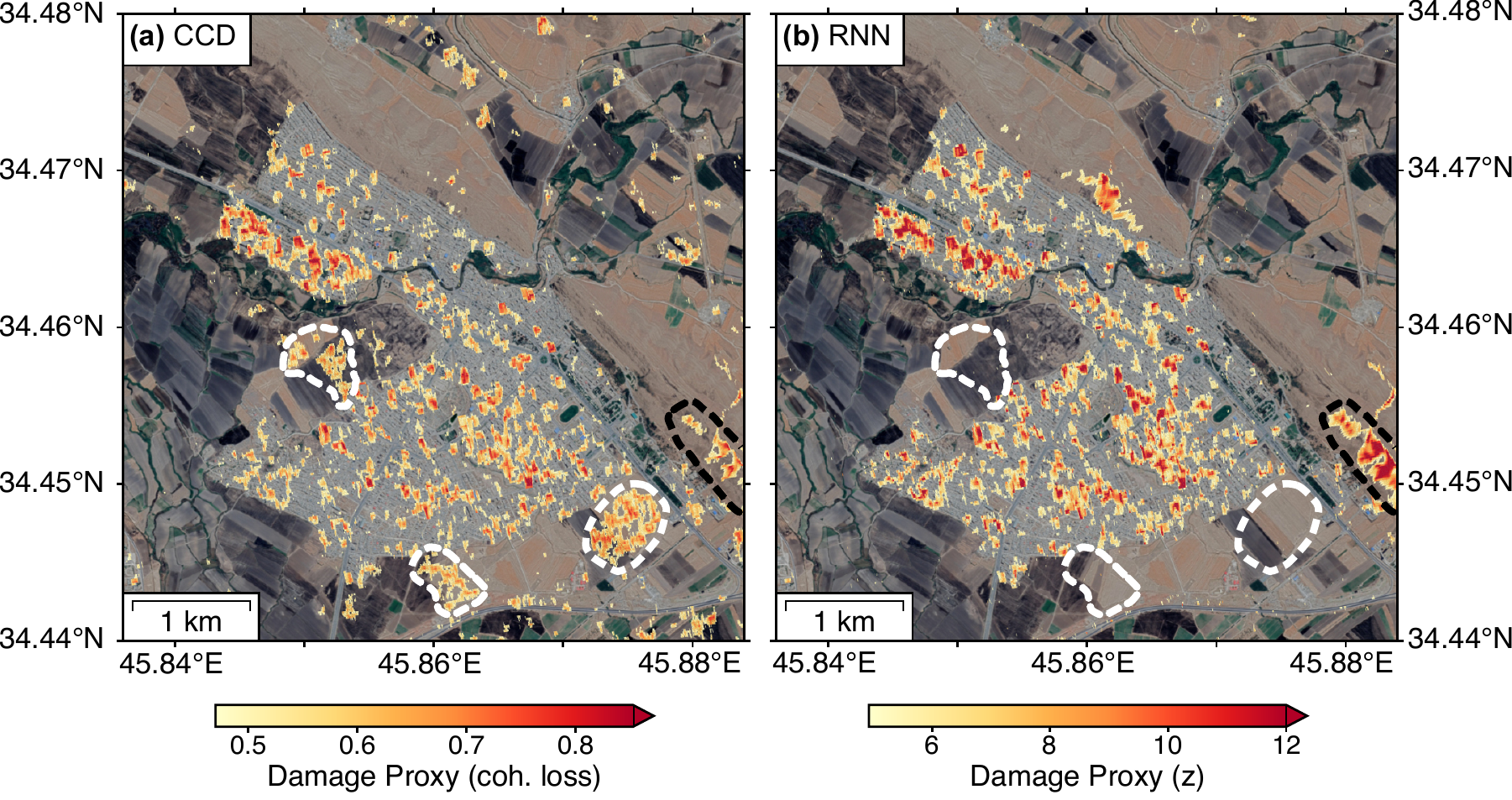}

\caption{Comparison of CCD \textbf{(a)} and RNN \textbf{(b)} DPMs for Sarpol-e-Zahab, damaged in the November 12, 2017 \Mw{7.3} Iran-Iraq earthquake. For each plot we mask values below a threshold, with the threshold chosen using the maximum value of the $\text{F}_{0.5}$ curve for the Amatrice data set (see Table \ref{table:opt_thresh}). The upper threshold of the color scales are chosen such that both plots have the same number of points above the threshold. The white dashed lines highlight example areas where the CCD method gives false positive damage detection in regions outside of the city that are no longer classified as damaged by the proposed RNN method. The black dashed line shows an area over a rocky ridge where greater damage is shown by the RNN method compared to the CCD method. Ground truth damage data is presented in Figure \ref{ext-fig:sarpol_ground}. Optical imagery from Google, CNES/Airbus, taken July 27th 2020.}
\label{fig:sarpol_compare}
\end{figure*}

In Fig. \ref{fig:sarpol_compare} we present damage proxy maps for the town of Sarpol-e-Zahab, damaged during the Iran-Iraq earthquake of November 12, 2017, for both the CCD and RNN methods. Using results from Amatrice (Table \ref{table:opt_thresh}) the threshold for damage is set at $z=4.93$ for the RNN method, and $\textrm{coherence loss}=0.47$ for the CCD method, with damage proxies below these thresholds masked out. For display purposes we again choose the upper limit of the RNN color bar to be $z=12$ and set the limit of the CCD color bar such that the same number of points are above the color bar limit as for the RNN case to provide a fair visual comparison. The damage data from the UN \cite{united_nations_institute_for_training_and_research_damaged_2017} (Figure \ref{ext-fig:sarpol_ground}) allows us to qualitatively compare the damage maps to the documented damage in the city.

Within the city, both methods highlight neighbourhoods where the UN located many collapsed buildings (for example in the north west of the city), however they also have elevated damage proxies over areas in the city where the UN did not record damage, possibly due to the sensitivity of InSAR coherence to small changes in surface properties, and possibly due to significant damage that was missed in the UN damage map. This seems to be more significant for the RNN method, which finds a larger amount of damage in the city than the CCD method. 

Looking outside the city yields a clearer difference between the methods. In Fig. \ref{fig:sarpol_compare} we use white dashed lines to highlight several areas where the CCD method has agricultural fields outside the city with high damage scores that are no longer highlighted in the RNN damage map, indicating that the co-event coherence for these areas was within the bounds of the normal variability for those pixels. This difference between the methods indicates the advantage of taking into account the full temporal behavior of each pixel. In Fig. \ref{fig:sarpol_compare} we also highlight one area outside the city where the RNN method has a higher damage proxy than the CCD method. As this area is over a rocky ridge, it is possible that the RNN damage proxy is capturing surface change due to rockfalls caused by the earthquake shaking. 

\begin{figure*}
\includegraphics[width=\textwidth]{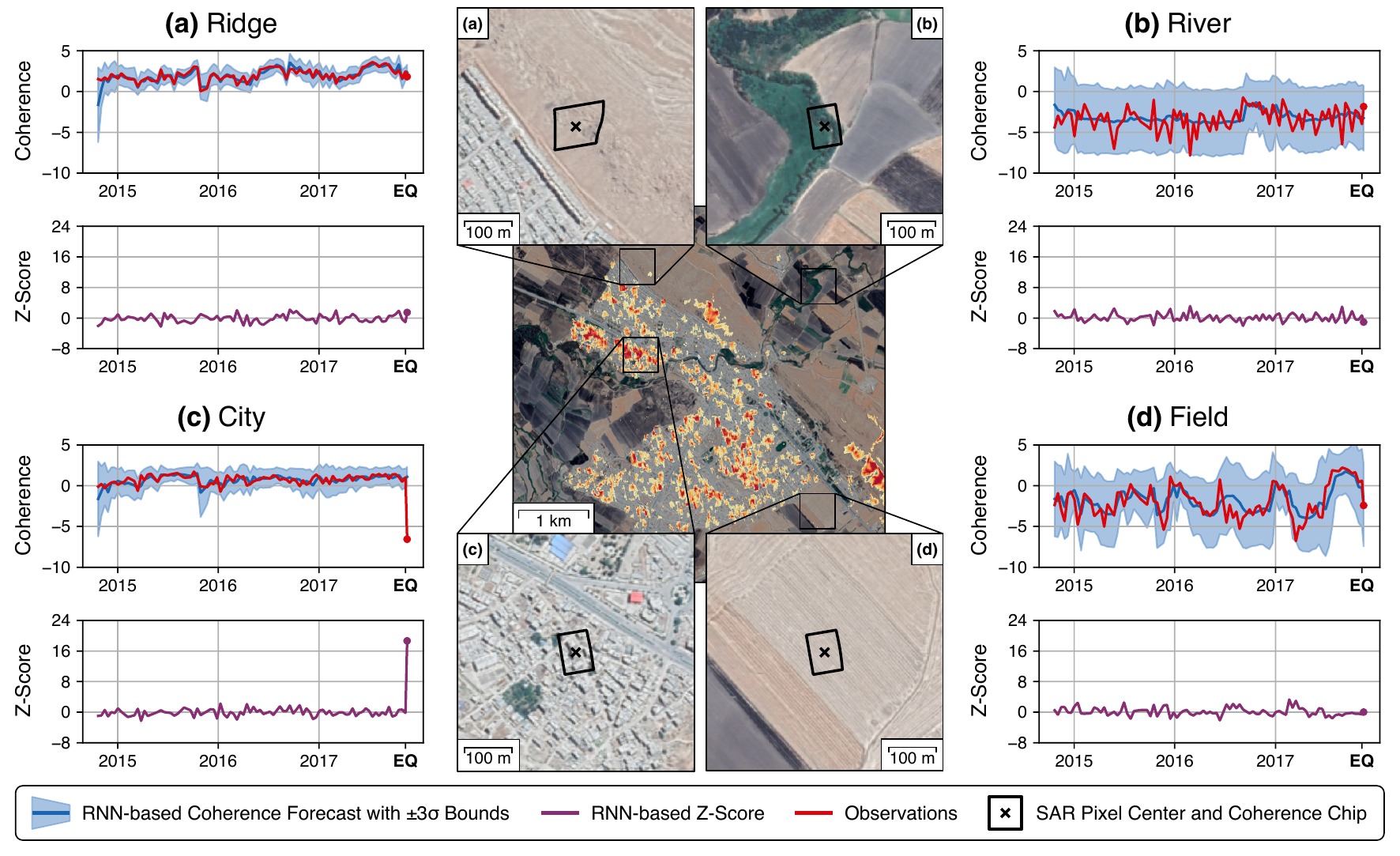}
\caption{Sarpol-e-Zahab RNN damage map along with coherence time series and Gaussian forecasts for four representative locations (\textbf{(a)}-\textbf{(d)}) around the city. The z-score indicates the number of standard deviations between the forecast and the ground truth, and the coherence is plotted in logit space (i.e. it has been transformed into an unbounded space, see Eq. \ref{eq:inverse-sigmoid}). Note that the shape of the coherence chip on the ground changes depending on the topography due to the way SAR data is acquired. The ``Coherence'' plotted on the y-axis is the logit transform of the squared coherence (Eq. \ref{eq:inverse-sigmoid}). Optical imagery from Google, CNES/Airbus, taken July 27th 2020.}
\label{fig:sarpol_time_series}
\end{figure*}

To better understand the damage map produced by the RNN, we select four locations in and around the town that show different styles of coherence time series. We apply this analysis to Sarpol-e-Zahab due to the larger amount of pre-event data compared to Amatrice, and the wider variety of pixel behaviors in a small area compared to Ridgecrest. In Fig. \ref{fig:sarpol_time_series}, we present the coherence time series, as well as the mean and standard deviation of the RNN forecast coherence and resulting z-score through time. For each case, the forecast at each time step is made based on Eq. \ref{eq:gauss_fcnn}, using the hidden state that is output from the trained model with the input being the coherence time series at that pixel up to that time. 

For pixel \textbf{(a)} of Fig. \ref{fig:sarpol_time_series}, over a rocky ridge, we see a high, stable coherence through time, with no substantial drop in coherence coseismically, hence a low co-event z-score. Pixel \textbf{(b)} is over a river, where surface properties change rapidly between SAR acquisitions, hence the pixel has a low coherence and higher uncertainty, but again has no co-event drop in coherence. Pixel \textbf{(c)} is within one of the badly damaged areas of the town. The pre-event coherence is high and relatively stable through time, causing a narrow uncertainty in the forecast. The co-event coherence is around 19 standard deviations below the forecast ($z \approx 19$), implying that the pixel is well out of the bounds of normal behaviour. We infer that this is due to building damage. Finally, pixel \textbf{(d)}, covering an agricultural field, has highly variable coherence through time, which causes a large variance in the forecast coherence. The co-event coherence is substantially below the final pre-event coherence, meaning the CCD method shows elevated damage proxy values, however in the context of the entire time series we see that this coherence is well within the bounds of the forecast coherence variability and thus the z-score is small.    

\subsection*{July 2019 Ridgecrest earthquakes, California, USA}
\begin{figure*}
\includegraphics[width=0.96\textwidth]{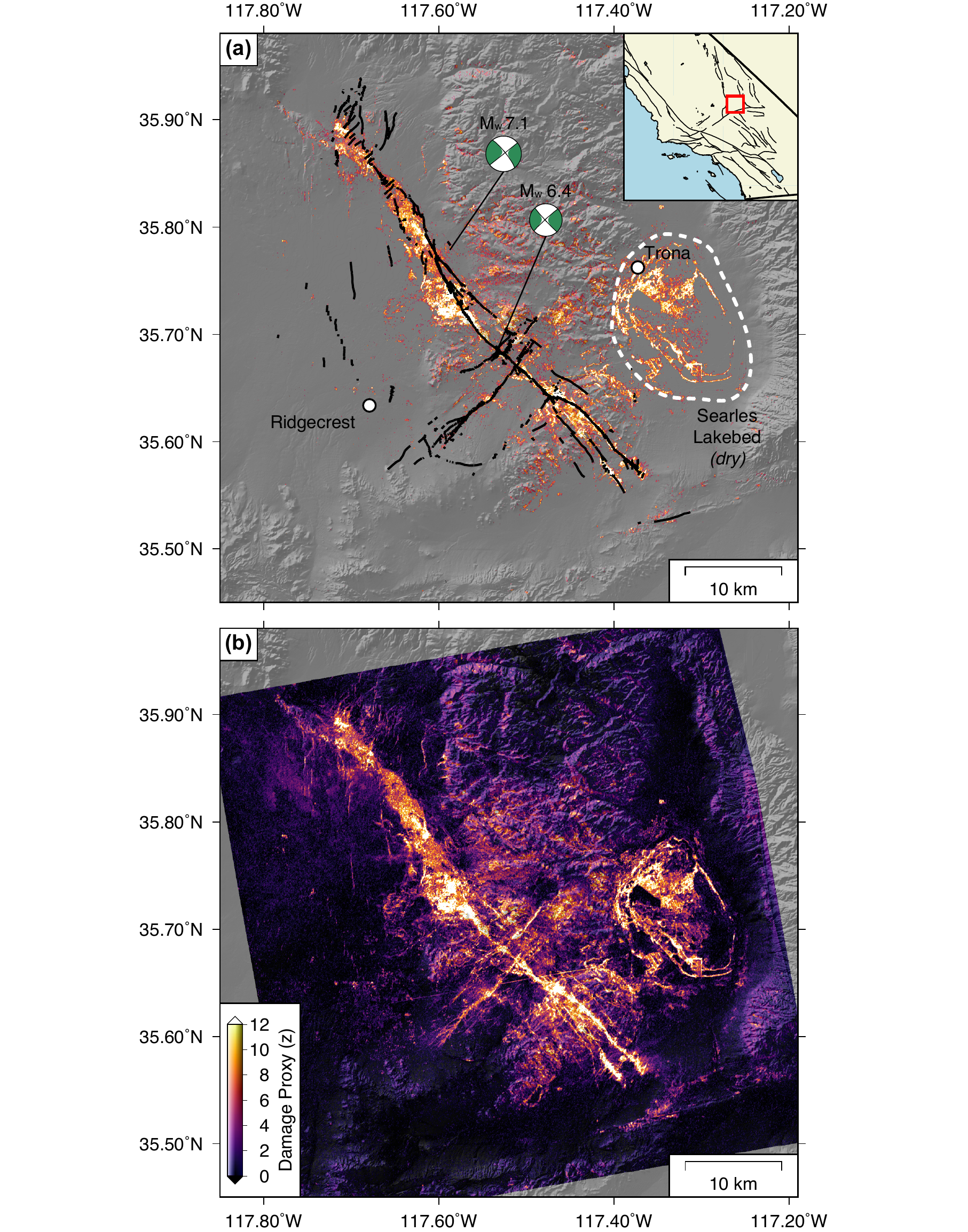}
\caption{Ridgecrest RNN damage proxy map with different thresholds. \textbf{(a)} All points with $z<4.93$ are masked. Black lines indicate mapped surface ruptures from Ponti \textit{et al.} \cite{ponti_documentation_2020}. The white dashed line indicates the approximate extent of the dry Searles Lakebed that saw substantial liquefaction \cite{zimmaro_liquefaction_2020}. Global CMT focal mechanisms are plotted for the \Mw{6.4} and \Mw{7.1} earthquakes \cite{dziewonski_determination_1981,ekstrom_global_2012}. The inset shows a regional map with simplified Quaternary faults. \textbf{(b)} Same damage proxy map as \textbf{(a)} except with no masking and points with $z<0$ plotted in black. This threshold allows less intense off-fault anomalous change to be more clearly seen.}
\label{fig:ridgecrest_dpms}
\end{figure*}

The earthquakes that struck near the town of Ridgecrest, California, in July 2019 \cite{ross_hierarchical_2019} provide an opportunity to test our proposed method on other forms of anomalous changes in ground surface properties, including fault surface rupture, landslides and liquefaction \cite{stewart_preliminary_2019, hough_nearfield_2020, kendrick_geologic_2019, ponti_documentation_2020,zimmaro_liquefaction_2020}. In Fig. \ref{fig:ridgecrest_dpms} we plot the RNN damage proxy map with two different z-score thresholds. The higher threshold allow us to focus on points that have had more anomalous coherence drops compared to their previous behaviour through time. Using a threshold of $z=4.93$ from the Amatrice data above (Fig. \ref{fig:ridgecrest_dpms} \textbf{(a)}) we see that the largest anomalies lie on the \Mw{7.1} and \Mw{6.4} surface ruptures (running NW-SE and NE-SW respectively) and liquefaction in the Searles Lakebed area, around (117.3\degree W, 35.6\degree N). 

In Fig. \ref{fig:ridgecrest_dpms} \textbf{(b)}) we plot all points below $z=0$ as black, more clearly showing smaller coherence anomalies surrounding the ruptures. Comparison with the mapped ruptures shows that some of these anomalies are due to smaller off fault ruptures, and we can also locate a small amount of damage in the town of Ridgecrest \cite{hough_nearfield_2020}. The correlation with topographic slope of many of the smaller anomalies (for example around 35.80\degree N, 117.50\degree W) suggest that these are due to small rockfalls or landslides induced by the earthquake. Damage maps such as these could be useful for directing mapping of ground failure in the aftermath of earthquakes. 

\section{Discussion}\label{s:discussion}

\subsection{Importance of the training data}

The goal of the RNN method is to produce the best possible forecast of the distribution of the co-event coherence value at each location, in the absence of any damage, given that location's pre-event coherence time series and the trained model. The forecast at every location depends on the trained model and thus contains information from every coherence time series used in the training. In this way, every forecast uses information learned from a wide spatial area. 

Different parts of a given geographic region will be affected by processes that affect coherence (e.g. rain and snow) in a similar fashion, meaning that some amount of correlation between coherence time series in the region is likely. For example, a storm could cause consistent amounts of change in the surface properties across a wide area, leading to a sudden, correlated drop in coherence for many of the time series in the region.

When training the RNN, we split the overall training set $\mathcal{D}_t$ into training ($\mathcal{D}_{t,t}$) and validation ($\mathcal{D}_{t,v}$) components. $\mathcal{D}_{t,t}$ is used to optimize the RNN parameters, whereas $\mathcal{D}_{t,v}$ is used to evaluate the model performance according to Eq. \ref{eq:loglikelihood} and choose the model with the best loss (note the ultimate task of damage classification is not part of model selection). When the training set $\mathcal{D}_t$ is split into its training and validation components, any correlation between time series from the same geographic area could lead to \textit{data leakage}\cite{kaufman_leakage_2011}, whereby information from the validation set can also be found in the training set.

Data leakage can result in the RNN making artificially good forecasts as the RNN memorizes correlated patterns in the data, effectively over-fitting rather than learning generalizable rules. For example, we might get an unexpectedly good forecast of a sudden coherence drop in a time series which the RNN had not previously seen, due to the same pattern also being present in time series used in training. A possible case of this can be seen in time series \textbf{(a)} of Fig. \ref{fig:sarpol_time_series}; in late 2015 there is a sudden drop in coherence (from around 3 to 0 in logit space) which is closely mirrored by the mean and variance of the forecast distribution. It seems unlikely that this coherence drop would be so accurately forecastable unless the network had seen many examples of similar patterns during training. Ideally, the network would instead learn that sudden drops in coherence can occur, and would broaden its uncertainty (i.e. the forecast standard deviation) accordingly. 

Since data leakage can lead to artificially high performance on supervised machine learning tasks \cite{kaufman_leakage_2011}, we ask if data leakage during RNN training could lead to artificially improved damage classification. In our case, there can be no leakage of the actual ground truth damage data, which is the target for our RNN-based classifier but is not used in training. There can only be leakage in correlated patterns in the pre-event coherence time series. As detailed above, data leakage could cause over-fitting, making the forecast overly confident, i.e. with a standard deviation ($\sigma'$) that is too small. An overly confident forecast will lead to a z-score that is larger for the same difference between observed coherence ($x$) and the mean of the forecast ($\mu'$), making the z-score noisier and meaning the likely result of data leakage is a worse performing damage classifier, not one with artificially improved performance.

The quality of the RNN forecast will depend on the training data which we use. In general, training data acquired over a shorter time span and over a smaller spatial area will not sample the full scope of representative coherence behaviour. We expect that more limited training data will cause the model to give a less representative forecast coherence distribution, as the network will see fewer examples of how coherence can vary in time and space. In Section \ref{ext-a:training_data} we explore how decreasing the time span of the training data affects the quality of the damage map for the Amatrice example, finding that the results are highly variable when less than a year of data is used. Based on these results we emphasize the importance of testing the RNN method on a wider variety of disasters, in different geographic regions, to ensure robust performance.  

A smaller geographic region is more likely to have strongly correlated coherence time series, leading to more significant data leakage problems and over-fitting during training. Drawing pixels from a very wide geographic area, or potentially many different geographic areas all over the Earth, could ameliorate this problem. A systematic exploration of the relationship between the input SAR data and the quality of the final damage map is beyond the scope of this paper.

Coherence images with longer temporal baselines will generally have lower coherence than images from a similar time period and region with a shorter temporal baseline \cite{zebker_decorrelation_1992}. Our current RNN training approach ignores the temporal baseline of the coherence images, effectively assuming that all coherence images have the same temporal baseline. However, our coherence time series have some variation in temporal baselines due to the varying acquisition frequency of \mbox{Sentinel-1} SAR data, with the repeat frequency tending to become more stable and more frequent with time (see Section \ref{ext-a:study_areas} for details of the data). A variable temporal baseline adds an extra noise term due to changing amounts of temporal decorrelation. Therefore, the likely effect of a variable temporal baseline is to make the coherence time series noisier, thus decreasing the confidence of the forecast and reducing the sensitivity to damage.

For the Amatrice case, the timing of the event meant that we had less than two years of \mbox{Sentinel-1} data preceding the earthquake, and the data had a higher variance in the temporal baseline. The lack of data and variable temporal baseline degrades the performance of the RNN, so the precision-recall results presented are likely a lower bound on the possible performance for damage mapping in this area. 

\subsection{Sensitivity to damage in different geographic areas}
Random motion of scatterers within a resolution element on the scale of the wavelength of the satellite radar signal ($\approx$5.6 $\si{cm}$ for this work) will cause decorrelation between two radar echoes \cite{zebker_decorrelation_1992}. Different regions of any study area will have different background rates of change in their surface properties, and these rates may vary through time. High rates of surface change will lead to low coherence, and variability through time in the rate of change will create a larger standard deviation in the coherence time series. 

Robust damage detection relies on separating normal changes from damage induced changes, and the ability to do this separation depends on the rate of surface change and how much this rate varies in time. At a given time step, for a given pixel, the RNN forecasts the average rate of change with the mean of the forecast ($\mu'$), and the variability with its standard deviation ($\sigma'$), both of which are used to calculate the z-score (Eq. \ref{eq:z_score}). A high background rate of surface change will lead to a low $\mu'$, thus obscuring coherence drops due to damage as $\mu'-x$ will be small. Similarly, large coherence variability will cause larger $\sigma'$ values leading to smaller z-scores, making it hard to separate coherence drops that are natural from those that are damage induced. 

Differing behavior of pixels means that, for two different pixels, the same z-score does not necessarily imply the same level of damage, but instead the same ratio of coherence change to background coherence variability. Thus, when interpreting the z-score map, it may also be useful to consider the forecast mean and standard deviation to understand the noise level for each pixel, with low mean and high standard deviation indicating noisy pixels and thus lower sensitivity to damage. We note that stable, human-made structures typically have higher and less variable coherence, whereas areas with vegetation, water and snow typically have lower and more variable coherence. 

For the Ridgecrest damage proxy map (Fig. \ref{fig:ridgecrest_dpms}), it is noticeable that the largest z-scores appear to correspond to the most significant ground disruption, specifically the surface ruptures from the \Mw{6.4} and \Mw{7.1} earthquakes and liquefaction near the town of Trona \cite{stewart_preliminary_2019,hough_nearfield_2020,zimmaro_liquefaction_2020}. The apparent link between z-score magnitude and intensity of ground surface change may be due to the desertic conditions in the area. The dry, stable conditions mean that most pixels have similar behavior through time, i.e. they have a similar noise level, meaning z-scores are more directly comparable between pixels. More generally, we should be able to use z-scores as proxies for levels of damage within groups of pixels that have similar forecasted standard deviations. However, the z-score is less comparable between groups of pixels with very different forecasted means and standard deviations.

The desertic conditions in the Ridgecrest area mean that coherence is comparatively high and stable through time. Because of this stability, the single pre-event image used in the CCD method is a better proxy for the pre-event coherence than the single images used in the other regions, causing the RNN and CCD methods to be more similar than for the other two case studies (we do not present the CCD results here).

Damaged buildings that are smaller or on poorly orientated slopes with respect to the satellite line of sight will occupy a smaller fraction of the coherence chip. These buildings will therefore have a smaller effect on the coherence, making them harder to identify using coherence based methods. The choice of the fraction of the chip area which has to be occupied by damaged buildings in order for that chip to be in the ``truly damaged'' class can therefore have a substantial effect on the final precision-recall area under curve (PR AUC) values for the Amatrice data set. We choose one third of the chip area (roughly 1200 $\si{m}^2$) as it gives approximately the largest PR AUC values, although the change in PR AUC values between fractions of 20\% and 50\% are small (RNN PR AUC in the range 0.68-0.70) The decrease of the RNN PR AUC below 20\% building fraction suggests that our method has difficulty identifying damage where the damaged area occupies less than around 700 $\si{m}^2$ of the coherence chip, meaning, for example, our method could have difficulties correctly classifying an isolated damaged building that is smaller than 25 $\si{m}$ $\times$ 25 $\si{m}$. With the present resolution, this method is likely most useful for detecting large damaged buildings and damaged blocks, rather than damage to individual smaller buildings. 

\subsection{The choice of change metric and forecast distribution}

While coherence has proven useful for surface change detection, metrics such as the SAR amplitude correlation can also be used \cite{jung_evaluation_2020}. Using the amplitude can be particularly useful when the InSAR coherence is low. We suspect that a similar RNN-based anomaly detection approach would also work on time series of other metrics used for SAR change detection, and could also be combined with the coherence metric, however we do not pursue this here.

When calculating coherence, we approximate an ensemble average for a given SLC pixel with a spatial average around that SLC pixel when evaluating Eq. \ref{eq:coherence} (e.g. see Section 3.12.2.5 of \cite{simons_interferometric_2015}). In our case, we use a local chip (15 by 5 SLC pixels in size) and assume that these SLC pixels are drawn from the same statistical distribution of amplitude and phase. There is a trade off in the choice of chip size—smaller chips are more likely to combine pixels that demonstrate the same statistics through time (as a result of being over the same type of land surface), however they also contain fewer samples with which to estimate the coherence, leading to a larger variance in the estimate and a noisier coherence time series. Larger coherence chips contain more samples, but are more likely to include pixels with greater differences in statistical behavior through time, less consistent with the assumption that we used to approximate the ensemble average. Larger chips also provide lower spatial resolution. Limited testing with the data presented in this study indicates that the results are made worse by reducing the chip size to 9 by 3 SLC pixels, however we have not systematically explored how the results vary with chip size.

The choice of forecasting a Gaussian distribution on the logit transform of coherence squared is motivated by the desire to use an unbounded distribution on an unbounded space, and the mathematical relationship to the logarithm of the phase variance (Section \ref{ext-a:phase_var}). However, the specific transform and distribution are ad-hoc and in general they may not produce the optimal forecast. We leave the exploration of the best transform and distribution to future work.

Rather than using a local chip to estimate coherence, a stack of SAR images can be used to identify a non-local neighborhood of SLC pixels, within some distance of a central SLC pixel, that behave in a statistically similar pattern through time. The coherence can then be evaluated over those SLC pixels, which has been shown to give a better coherence estimate \cite{ferretti_new_2011}. However, using a non-local coherence evaluation introduces problems when doing damage mapping. Damage will change the statistical characteristics of an SLC pixel, meaning that co-event SLC pixels may no longer be in their pre-event statistical groupings. For example, some of the SLC pixels may be over collapsed buildings, with other SLC pixels over buildings that remained undamaged. On the other hand, SLC pixels in a local chip are more likely to have been affected by the same process (e.g. building collapse). Thus, while a non-local coherence calculation may give improved results for pre-event coherence calculations, our method requires us to use a local chip for the coherence calculation. 

\subsection{Near real-time deployment}

When deploying damage mapping for rapid response, delivering timely products, ideally within hours, is exceptionally important. While the wait time for a \mbox{Sentinel-1} post-event acquisition could be up to six days, in many cases we will have an image before this, allowing the information to feed in to rapid disaster response. Other SAR satellites also acquire data, however they do not have the same long time series of open access acquisitions that is available from \mbox{Sentinel-1}. Planned SAR missions, such as NISAR \cite{sharma_updates_2019} and ALOS-4 \cite{motohka_alos-4_2019} should improve the availability of frequently acquired SAR data.

A near real-time deployment could follow this workflow, which would ideally be almost completely automated:
\begin{enumerate}
    \item Identify the natural disaster
    \item Find the SAR satellite which has a sufficient time series of existing observations, and will have an overflight of the affected area in the immediate future
    \item Download the existing SAR data, coregister, and calculate a coherence time series
    \item Train the RNN and forecast the co-event coherence
    \item Obtain the first post-event acquisition, coregister and calculate co-event coherence
    \item Compare with the forecast to calculate the damage proxy map
    \item Inspect and distribute to first responders via previously established channels of communication, ensuring that they have a clear understanding of the information the damage proxy map is providing.
\end{enumerate}

The most computationally demanding steps of this process is the coregistration of the large quantities of pre-event SAR data, with the coherence time series calculation and RNN training also requiring substantial computational resources. Combined, these steps can take several days of computing using our current codes and resources, potentially impacting response times. However, steps 1--4 do not necessarily affect the post-disaster response time, as they can either be pre-computed and regularly updated, or in some cases, be completed before the essential post-event acquisition becomes available. Additionally, the use of cloud computing resources and improved algorithms can greatly decrease processing time. Free and open data, accessible with minimal latency, are vital for the effective deployment of such a disaster monitoring system. 

The quality of the damage map could be further enhanced by combination with other damage assessments, e.g. maps of shaking intensity and on-the-ground reports, as well as previously identified zones of higher risk for building collapse, fault surface rupture, landslides and liquefaction \cite{loos_g-dif_2020}.

\section{Conclusion and Future Work}\label{s:conclusion}
In this work we present a deep learning-based damage mapping algorithm for synthetic aperture radar (SAR) sequential coherence time series. Coherence represents a proxy for ground surface change, and thus by separating out anomalous from expected coherence values after a natural disaster we can find regions that have had anomalous surface change, potentially due to building collapse, surface rupture, landslides or other hazards. We use a recurrent neural network to learn the normal behaviour of coherence through time by training on SAR coherence time series spanning a large area, then forecast the probability distribution of the coherence we expect without any disaster. We then use the deviation between the observed and forecast coherence to locate anomalous coherence changes, which we assume to be due to collapsed buildings. A comparison with on-the-ground building damage assessments shows that this method is quantitatively better than an alternative method of damage mapping based on coherence loss. We discuss the advantages and limitations of our proposed SAR damage mapping method and outline how it could be deployed in disaster response scenarios. 

The problem of RNN over-fitting due to the correlated nature of coherence time series in a single region could potentially be ameliorated by simultaneously training on a large number of coherence time series drawn from areas all over the planet displaying very different temporal behaviors. Furthermore, including additional training features such as the spatial and temporal baseline and the amount of precipitation between sequential SAR acquisitions might allow the network to learn the dependence on additional physical parameters relevant to coherence, thus improving its forecast. The ability to learn from many different input features without a physical model is a key advantage of the deep learning approach. 

The work presented here has been using C-band (5.6 $\si{cm}$ wavelength) SAR data. As coherence is sensitive to surface disruptions on the scale of the radar wavelength, it could be that we are picking up many false positives caused by superficial damage. Investigating the same disasters with 24 $\si{cm}$ wavelength L-band data may provide damage maps that are less prone to pick up small surface disturbances. Unfortunately, dense time series of L-band SAR data are not easily available, although the future launches of L-band SAR satellites (e.g. NISAR \cite{sharma_updates_2019} and ALOS-4 \cite{motohka_alos-4_2019}) will allow for further exploitation of L-band SAR data for damage mapping. NISAR will also image selected regions using S-band radar, allowing the use of SAR data at multiple wavelengths to further increase the ability of the damage map to distinguish different types of surface change.

\section*{Acknowledgments}
SLC processing was performed using the InSAR Scientific Computing Environment (ISCE) software from JPL/Caltech \cite{rosen_insar_2012}. Part of this research was supported by the National Aeronautics and Space Administration Applied Sciences Disasters Program and performed at the Jet Propulsion Laboratory, California Institute of Technology. This work contains modified Copernicus data from the Sentinel‐1A and 1B satellites processed by the European Space Agency and downloaded from the Alaska Satellite Facility Distributed Active Archive Center. Plots were produced using Matplotlib \cite{hunter_matplotlib_2007}, Cartopy \cite{met_office_cartopy_2010} and GMT \cite{wessel_generic_2013}. The RNN was built and trained using PyTorch \cite{paszke_pytorch_2019}. SRTM version 3 was used for the Ridgecrest digital elevation model. The authors would like to thank Heresh Fattahi and Piyush Agram for producing the coregistered SLC stack for the Sarpol-e-Zahab region, and Eric Fielding for producing the Ridgecrest stack. The authors would also like to thank Prof. Richard Murray, Stuart Feldman and Eric Schmidt for assisting with the early development of the work. The authors are also grateful to Prof. Mahdi Motagh and two anonymous reviewers for helpful comments on the manuscript.


%


\bibliographystyle{IEEEtran}
\bibliography{main}

%
%
\begin{IEEEbiography}[{\includegraphics[width=1in,height=1.25in,clip,keepaspectratio]{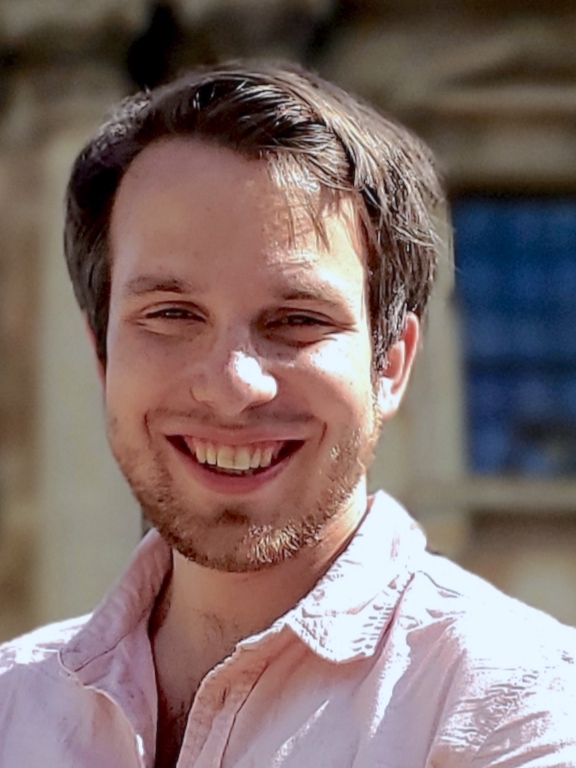}}]{Oliver L. Stephenson}
received his B.A. and M.Sci. (2015) in Natural Sciences (Physics) from Cambridge University, UK, and his M.Sci. (2019) in geophysics from the California Institute of Technology, Pasadena, CA, USA. His research interests include studying earthquakes using remote sensing methods and computational simulation. He is currently a Ph.D. candidate in Geophysics in the Seismological Laboratory, California Institute of Technology. (Photo credit: Dr. Susie Wright)  
\end{IEEEbiography}

\begin{IEEEbiography}[{\includegraphics[width=1in,height=1.25in,clip,keepaspectratio]{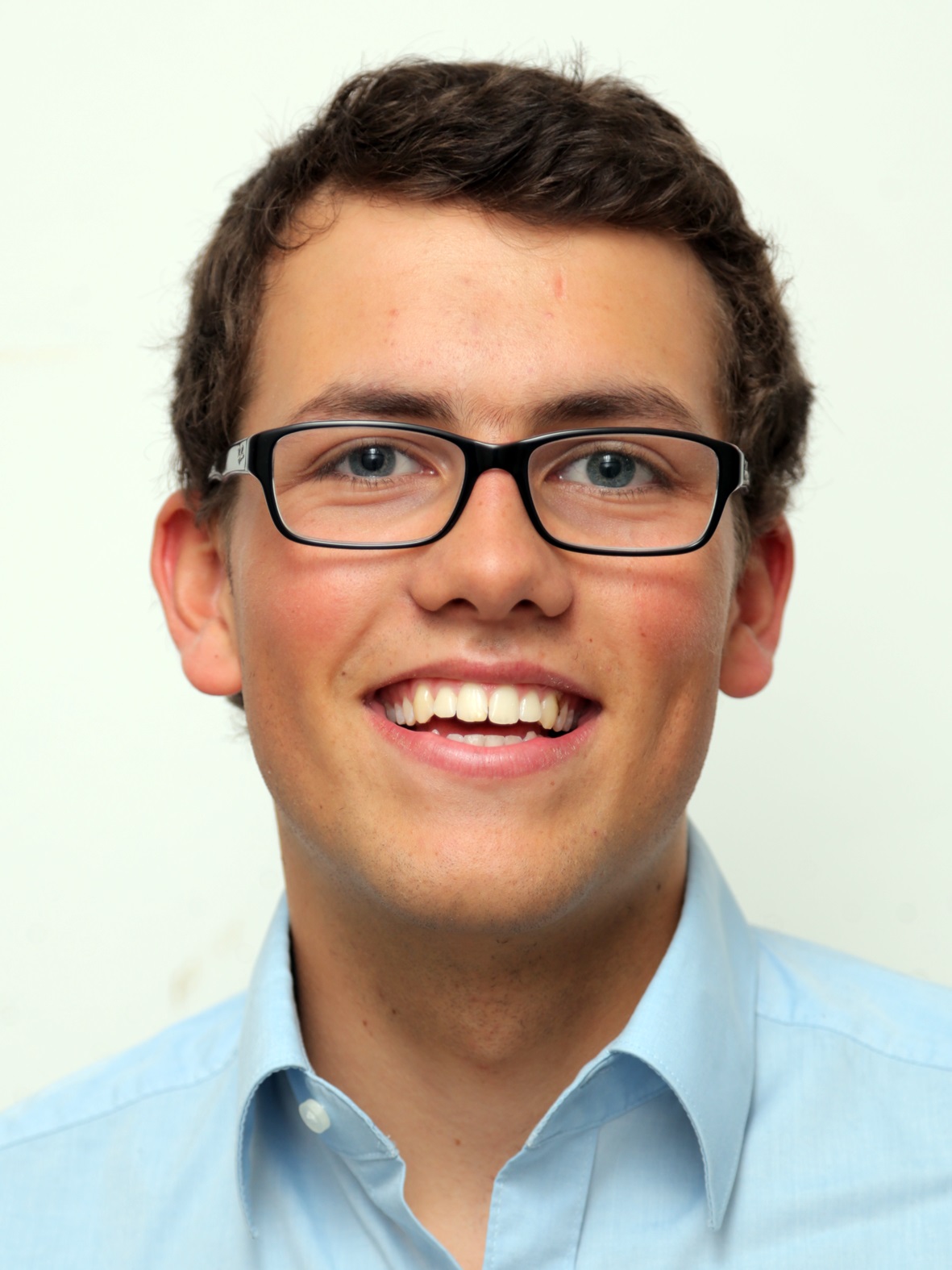}}]{Tobias Köhne} received his B.S. (2016) in Mechanical Engineering from the Technical University of Munich and his M.S. (2018) in Aerospace Engineering from the University of Texas at Austin focusing on orbital mechanics. He is currently a Ph.D. student at the California Institute of Technology, where he is working on modern tools for geodetic time series analysis of large-scale, continuously-operating GNSS regional networks. He is also part of an ongoing investigation into how machine learning approaches can help to extract more information out of currently available InSAR time series data sets.
\end{IEEEbiography}


\begin{IEEEbiography}[{\includegraphics[width=1in,height=1.25in,clip,keepaspectratio]{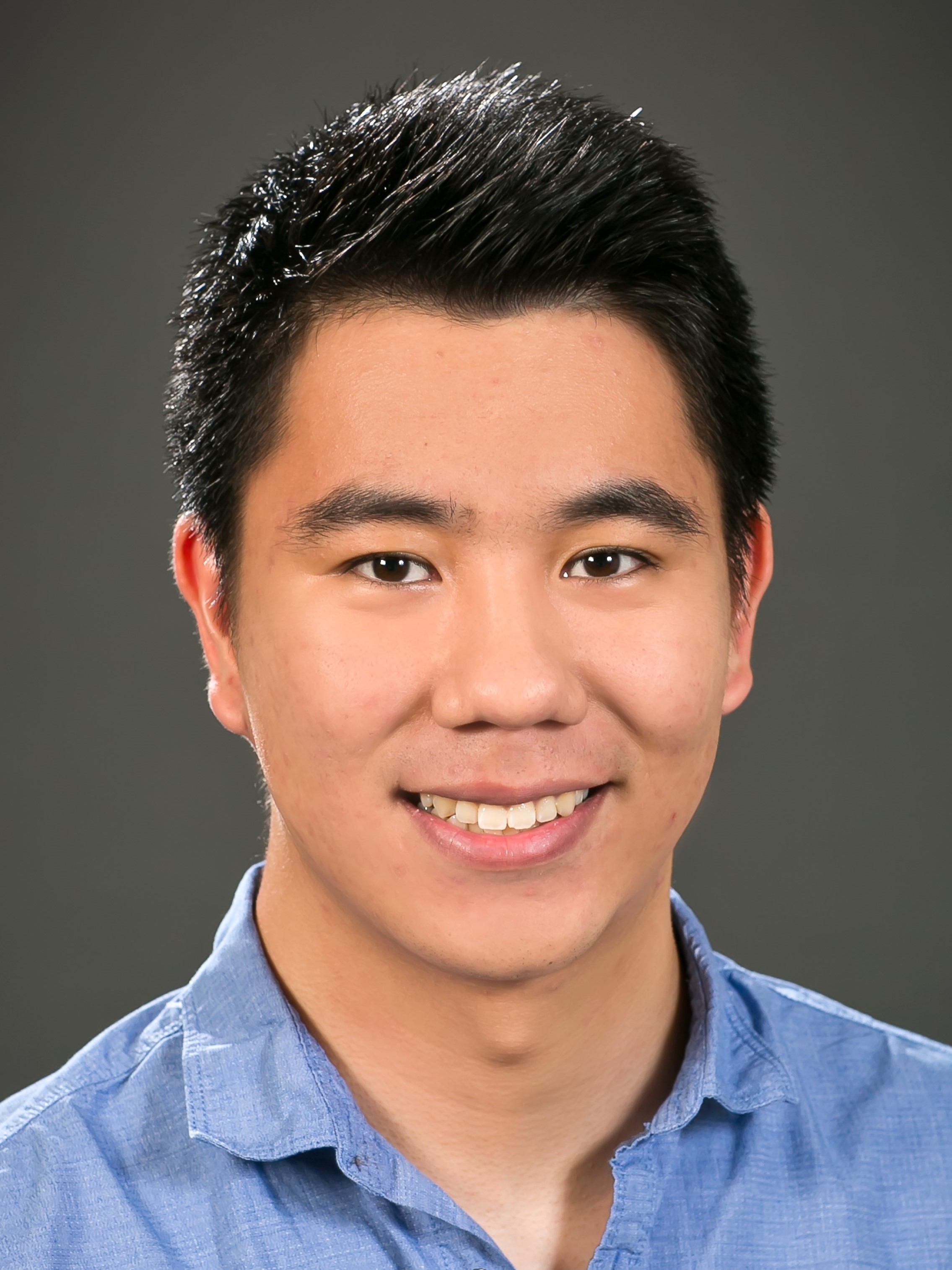}}]{Eric Zhan} received his B.A. (2016) in Computer Science and Mathematics (double major) from Cornell University. He is currently a Ph.D. candidate in Computing and Mathematical Sciences at the California Institute of Technology. His research interests include generative behavior modeling and sequential decision making, focusing on developing new methods for imitation learning inspired by generative models, weak supervision, and program learning.
\end{IEEEbiography}

\begin{IEEEbiography}[{\includegraphics[width=1in,height=1.25in,clip,keepaspectratio]{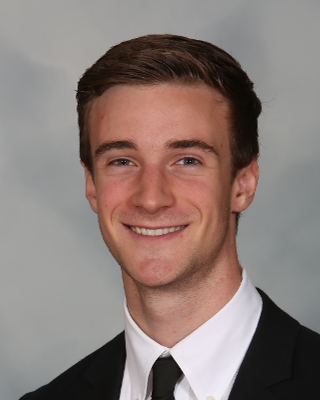}}]{Brent E. Cahill}
received his B.S. (2019) in Computer Science with concentration in machine learning and data science from the California Institute of Technology, Pasadena, CA, USA. While there, his research interests included deep learning-based natural disaster detection and damage mapping and machine learning in the analysis of the gut microbiome in patients with metabolic syndrome. He joined REMY Biosciences in Irvine, CA in 2019, where he sits as senior software and engineering manager, overseeing the research, development and production of alternative drug-delivery systems via transdermal and sublingual pathways. His main focus is the creation of data-driven approaches to precise systemic and local delivery of various neutra- and pharmaceuticals via clinical research.
\end{IEEEbiography}

\begin{IEEEbiography}[{\includegraphics[width=1in,height=1.25in,clip,keepaspectratio]{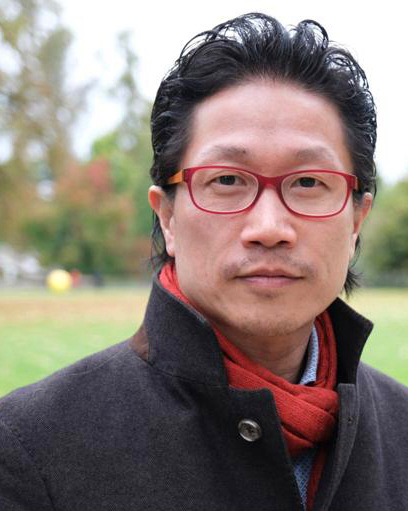}}]{Sang-Ho Yun} is a Geophysicist and Radar Scientist at the NASA Jet Propulsion Laboratory (JPL). He received his Ph.D. in Geophysics and Master’s in Electrical Engineering from Stanford University, and prior to joining JPL, he was a Postdoc at the U.S. Geological Survey (USGS). As the Disaster Response Lead of the Advanced Rapid Imaging and Analysis (ARIA) project at JPL and California Institute of Technology, he and his team have supported over 90 major disaster events globally. He received the 2018 NASA Exceptional Public Achievement Medal and the 2014 NASA Exceptional Early Career Medal for innovative use of SAR data in support of rapid post-disaster response. He is currently the Principal Investigator for three NASA projects to develop algorithms and systems for global rapid damage and flood mapping as well as human activity monitoring for the COVID-19 Global Pandemic response.
\end{IEEEbiography}

\begin{IEEEbiography}[{\includegraphics[width=1in,height=1.25in,clip,keepaspectratio]{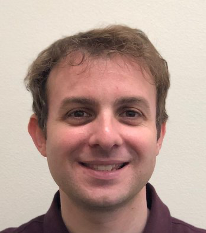}}]{Zachary E. Ross}
is an Assistant Professor of geophysics at California Institute of Technology (Caltech), Pasadena, CA, USA, where he uses machine learning and signal processing techniques to better understand earthquakes and fault zones. He is interested in seismicity, earthquake source properties, and fault zone imaging.
\end{IEEEbiography}
\enlargethispage{-5in}
\newpage

\begin{IEEEbiography}[{\includegraphics[width=1in,height=1.25in,clip,keepaspectratio]{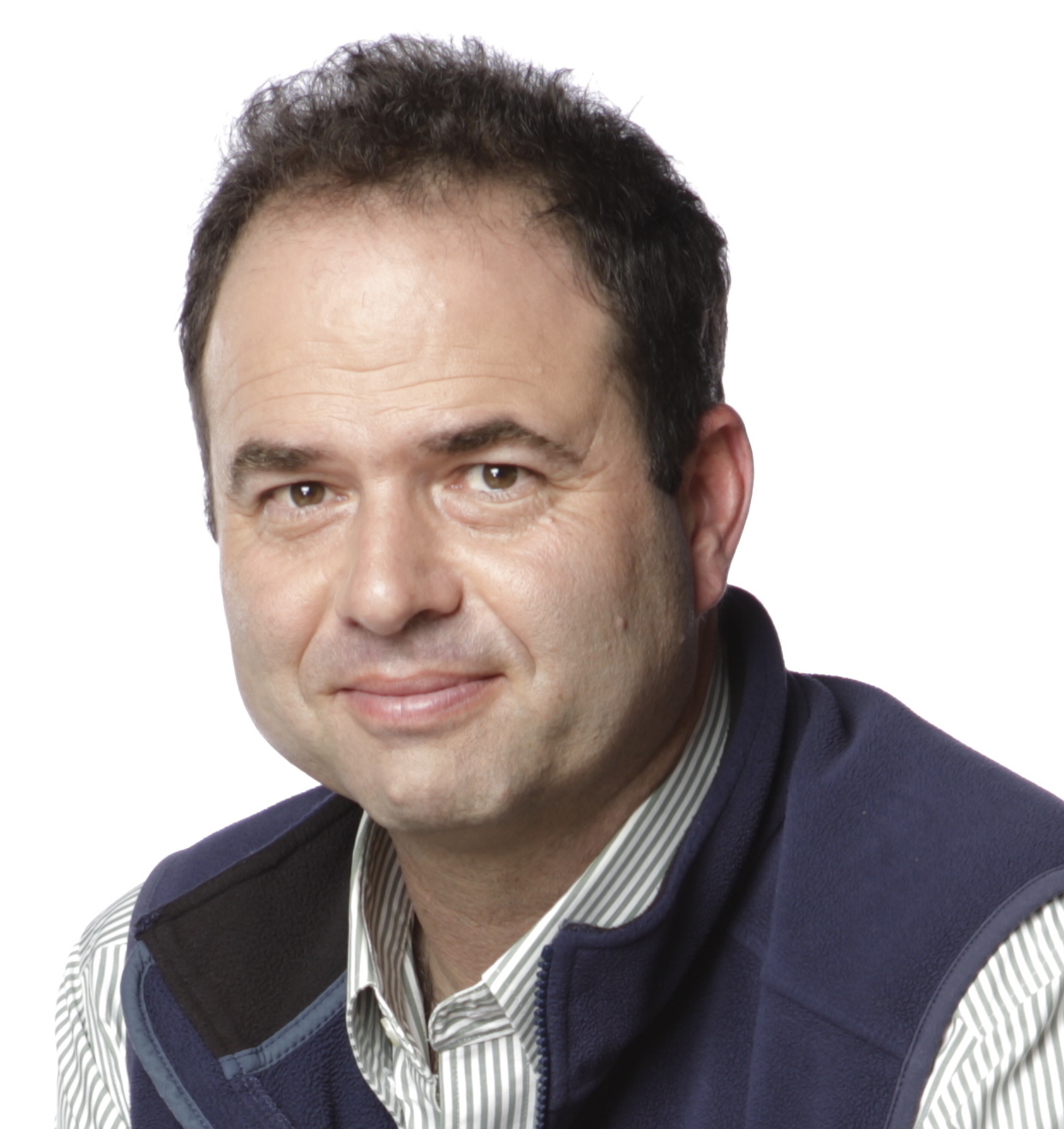}}]{Mark Simons} received the B.Sc. degree in geophysics and space physics from the University of California at Los Angeles, Los Angeles, CA, USA, in 1989, and the Ph.D. degree in geophysics from the Massachusetts Institute of Technology, Cambridge, MA, USA, in 1996. He has been with the California Institute of Technology, Pasadena, CA, USA, since 1996, where he is currently the John W. and Herberta M. Miles Professor of Geophysics at the Seismological Laboratory, Division of Geological and Planetary Science. He is also the Chief Scientist of the Jet Propulsion Laboratory. His research interests include processes associated with the seismic cycle, migration of magma and water in the subsurface, tides, and glacial rebound; tectonics and the relationship between short and long time-scale processes; glaciology, particularly basal mechanics and ice rheology; tools and applications using space geodesy, particularly GNSS and SAR; Bayesian methods for large geophysical inverse problems; and application of space geodesy for monitoring and rapid response to natural disasters. Prof. Simons is a fellow of the American Geophysical Union.
\end{IEEEbiography}

\ifCLASSOPTIONcaptionsoff
  \newpage
\fi




%


\end{document}


\title{Supplementary Materials for ``Deep Learning-based Damage Mapping with InSAR Coherence Time Series''}
%
%
%

\author{Oliver~L.~Stephenson,
        Tobias~Köhne,
        Eric~Zhan,
        Brent~E.~Cahill,
        Sang-Ho~Yun,
        Zachary~E.~Ross
        and~Mark~Simons}

%
%

\markboth{PAPER ACCEPTED FOR PUBLICATION IN IEEE TRANSACTIONS ON GEOSCIENCE AND REMOTE SENSING}{}%
%




\maketitle

\renewcommand{\thepage}{S\arabic{page}}
\setcounter{page}{1}

\renewcommand{\thesection}{S.\Roman{section}}
\renewcommand{\thesubsection}{\thesection.\Alph{subsection}}
\setcounter{section}{0}


\renewcommand\thefigure{S\arabic{figure}}  
\setcounter{figure}{0}


 \renewcommand\thetable{S\arabic{table}}  
 \setcounter{table}{0}

\renewcommand{\theequation}{S\arabic{equation}}
\setcounter{equation}{0}

\makeatletter
\renewcommand\@bibitem[1]{\item\if@filesw \immediate\write\@auxout
    {\string\bibcite{#1}{S\the\value{\@listctr}}}\fi\ignorespaces}
\def\@biblabel#1{[S#1]}
\makeatother

\section{Phase variance and the logit transform}\label{a:phase_var}

The Cramer-Rao bound on the variance of the interferometric phase, $\sigma^2_{phase}$, can be written \cite{rodriguez_theory_1992}:
\begin{equation}
    \sigma^2_{phase} = \frac{1}{2N_L}\frac{1-\gamma^2}{\gamma^2},
\end{equation}
where $\gamma$ is the coherence between two SLCs and $N_L$ is the number of SLC pixels in the chip used to estimate the coherence. The phase variance asymptotically approaches this limit as the number of looks increases, with the limit being a good approximation for $N_L > 4$. In this work our chip size gives us $N_L = 75$.

Taking the logarithm of the phase variance gives us:
\begin{equation}
    \log_e(\sigma^2_{phase}) = \log_e\left(\frac{1-\gamma^2}{\gamma^2}\right)-\log_e(2N_L).
\end{equation}
We also have the logit transform of coherence squared (Eq. \ref{ext-eq:inverse-sigmoid}), which is:
\begin{equation}
    S^{-1}(\gamma^2) = \log_e\left(\frac{\gamma^2}{1 - \gamma^2}\right).
\end{equation}
We therefore have the relationship:
\begin{equation}
    S^{-1}(\gamma^2) = -\log_e(\sigma^2) -\log_e(2N_L),
\end{equation}
meaning the logit transform of the coherence squared only differs from the logarithm of the phase variance by an additive constant and sign.

\section{RNN Model and Training Details}\label{a:rnn_details}
The basis of RNNs are \textit{artificial neural networks}. These are layers of matrix-vector multiplications, interspersed with non-linear \textit{activation functions}, that transform input data into a desired output. The matrix values (known as \textit{weights}) are varied, or \textit{trained}, to get the desired mapping between input and output data. For a quick pedagogic introduction to artificial neural networks, we refer the reader to \cite{3blue1brown_but_2017}, more detail on artificial neural networks and RNNs can be found in \cite{goodfellow_deep_2016}.

So-called \textit{vanilla} (basic) RNNs have trouble learning long-term dependencies, which in our case means learning how to make forecasts when the desired output at a given step depends on data received many steps in the past \cite{bengio_learning_1994}. Long short-term memory RNNs (LSTMs) were created to solve the problem of long-term dependencies \cite{hochreiter_long_1997}. LSTMs have a series of \textit{gates}—learned weights combined with non-linear activation functions—that allow the network to decide what information to add, retain, and forget when each new observation is fed in, in such a way as to better store information over many time steps compared to a vanilla RNN. For a pedagogic introduction to LSTMs, we refer the reader to \cite{olah_understanding_2015}. The weights are learned during training, so that the network is able to decide what information is most useful for a given set of time series. A single set of weights are learned for all the time series that are given to the network.

Gated recurrent units (GRUs) are a gating mechanism for RNNs, similar to LSTMs but with fewer parameters and thus less computationally expensive to train. GRUs have been found to have similar performance to LSTMs on a range of tasks \cite{cho_properties_2014}, and are what we use in this work.

As an example for our data, one could imagine trying to forecast the coherence for a region with dry summers, and wet winters. During the dry summer, the ground surface may be relatively undisturbed, leading to high InSAR coherence. However, rain and snow during the wet winters could  often disturb the Earth's surface, leading to large drops in coherence. If we had at least one year of coherence time series data, and were trying to forecast the coherence at the beginning of the summer, the recent coherence measurements from the winter months may not be the most useful data to make this forecast. Instead, we would want the network to make a forecast based on similar sequences that it had see during previous summer periods. 

By training the network on a large number of coherence time series containing winter and summer behavior, the gating mechanisms should learn the relevant information to use when forecasting during summer and winter sequences. Note that this is not simply learning a periodicity in the signal. If there was an unusually wet summer or dry winter, the network should be able to adjust its forecast based on new data and its training.   

\begin{figure*}
\centering
\includegraphics[width=\textwidth]{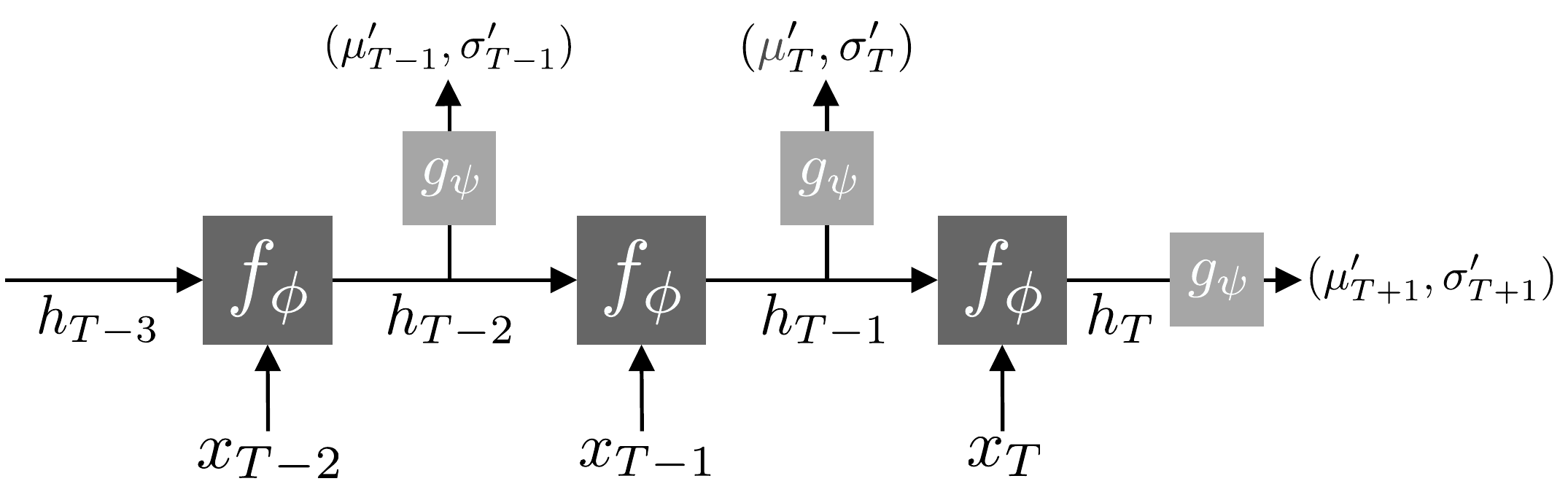}
\caption{\label{fig:graphical_model}Graphical model for our recurrent neural network for a series of time steps. We use gated recurrent units for $f_{\phi}$ and a feed-forward neural network for $g_{\psi}$. For each pixel the hidden state ($h$) summarizes the coherence information ($x$) up to that point, and is used to forecast the mean $(\mu')$ and standard deviation $(\sigma')$ of the logit-transformed squared coherence. We initialize the hidden state $h_0$ as all zeros. The final hidden state, $h_T$, is used to forecast the co-event coherence distribution parameters $(\mu'_{T+1},\sigma'_{T+1})$. }
\end{figure*}

The following equations describe $f_{\phi}$ (Eq. \ref{ext-eq:hidden_state}) for the GRU that we use in this work:

\begin{subequations}
\begin{IEEEeqnarray}{CC}
\IEEEyesnumber
    & z_t = \: S(W_z x_t + U_z h_{t-1} + b_z) \\
    & r_t = \: S(W_r x_t + U_r h_{t-1} + b_r) \\
    & \bar{h}_t = \: \text{tanh}(W_h x_t + U_h(r_t \circ h_{t-1}) + b_h) \\
    & h_t = \: (1 - z_t) \circ h_{t-1} + z_t \circ \bar{h}_t,
\end{IEEEeqnarray}
\end{subequations}
where $S$ is the sigmoid activation function, $\circ$ is the Hadamard (element-wise) product, $z_t$ is the update gate vector, $r_t$ is the reset gate vector, $\bar{h}_t$ is the candidate hidden state, and $\phi = [W, U ,b]$ are the learnable parameters of $f$. We use a hidden state of size 256, with the initial state $h_0$ set to all zeros. For $g_\psi$ (Eq. \ref{ext-eq:gauss_fcnn}), we use a fully connected, feed-forward neural network to map the RNN hidden state to parameters of a Gaussian distribution. This neural network has 3-layers, with Rectified Linear Unit (ReLU) activation functions and hidden layer sizes of 128. The resulting output is finally passed through two separate linear layers to obtain the mean and natural logarithm of the variance (we output the logarithm for improved numerical stability). Figure \ref{fig:graphical_model} shows how the different components are connected. In total, our model has 265,090 learnable parameters. Hyperparmeters of the architecture are chosen according to rules of thumb, and at this stage we have not performed a cross-validation or hyperparameter search. This search can be done to further improve the forecast made by the RNN, for example by varying the size of the hidden state and the number of layers in the fully connected neural network.

We randomly select 80\% of sequences in the training set $\mathcal{D}_t$ to train our model ($\mathcal{D}_{t,t}$) and use the remaining 20\% as validation sequences for model selection ($\mathcal{D}_{t,v}$). We emphasize that the model does not see any of the validation sequences and only sees pre-event information during training.

We train for 20 epochs with a batch size of 256 using the Adam optimizer with a learning rate of 0.0005 \cite{kingma_adam_2014}. Each epoch sees each training sequence once and has a total run time of several minutes on a Tesla P100 GPU (depending on the number and length of training sequences, which varies by region). The trained network is then applied to the forecasting data set $\mathcal{D}_f$ to generate co-event coherence forecasts at each time step and location. Details of how $\mathcal{D}_t$ and $\mathcal{D}_f$ are calculated are included in Section \ref{a:study_areas}.  

Non-determinism can be introduced in the RNN training, for example due to the implementation of GPU algorithms and the random selection of batches during training. From limited trials on the Amatrice data set we find that repeated training with the same data and training parameters can lead to small differences in the PR AUC (less than 0.05), but this variability does not affect our overall conclusions. It is possible that the data leakage issue, discussed in Section \ref{ext-s:discussion}, is leading to overfitting, and thus to a larger variance in the co-event forecast for repeated re-training, thus causing a larger variation in the PR AUC. We therefore caution that the specific values indicated in this paper shouldn't be taken as representative of the values likely to be obtained in all circumstances, and reiterate that further testing in a wider variety of cases is necessary. 

An implementation of our deep learning model can be found on GitHub: \url{https://github.com/olliestephenson/dpm-rnn-public}.

\section{Study Areas and Data}\label{a:study_areas}
A summary of the SAR data used for each study area is presented in Table 
\ref{table:study_areas}, and more details are given in this section. In all cases we use data starting in October 2014 (when the \mbox{Sentinel-1A} satellite began acquiring data) until the first post-event SAR acquisition. We start with Level-1 Single Look Complex (SLC) images acquired in interferometric wideswath mode \cite{european_space_agency_single_nodate}, then, using the InSAR Scientific Computing Environment (ISCE) \cite{rosen_insar_2012}, we create a coregistered stack of SLC images, which are corrected for the flat Earth and topographic phase contributions. For each region, we calculate $N-1$ sequential coherence images from $N$ SAR acquisitions. The post-event and final pre-event SAR acquisitions are used to generate a co-event coherence image, which is not seen during RNN training but is used to calculate the size of the coherence anomaly. While the time between SAR acquisitions varied for each region, in all three cases the first post-event image was taken six days after the final pre-event image. Note that we keep the data in radar coordinates for the coherence calculation and do not interpolate or resample once the SLC stack is coregistered.   

\begin{figure}
    \centering
    \includegraphics[width=0.48\textwidth]{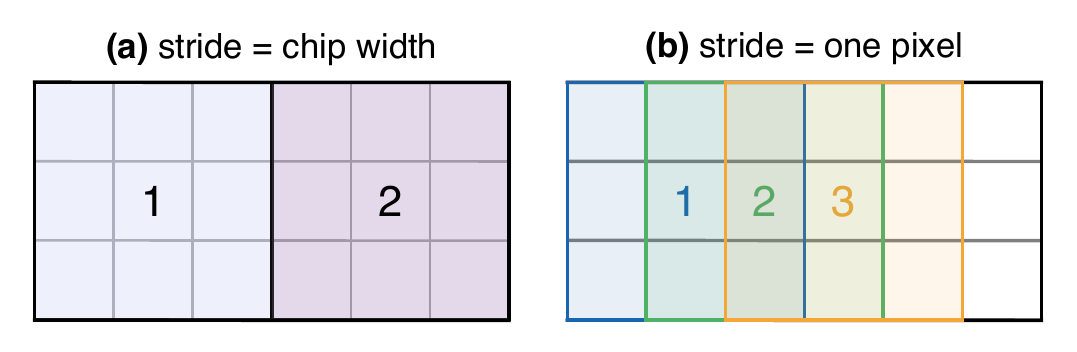}
    \caption{Schematic of the different coherence calculation methods for \textbf{(a)} $\mathcal{D}_t$, and \textbf{(b)} $\mathcal{D}_f$. The small squares represent SLC pixels and the large, numbered, squares represent coherence chips, here represented as three by three SLC pixels for illustration purposes.}
    \label{fig:chip_stride}
\end{figure}

To calculate the coherence time series in $\mathcal{D}_t$, we set the stride (distance between adjacent center SLC pixels of the chip, or averaging window) to be equal to the chip width in each direction so that each SLC pixel is only used to estimate coherence in one chip (Fig. \ref{fig:chip_stride}\textbf{(a)}). We use data spanning a wide area surrounding the specific region we are searching for damage.  

For calculating $\mathcal{D}_f$, we only use SAR data in the area of interest (i.e. the damaged region we would like to map), but set the stride of the coherence chip to one, giving us a larger number of coherence pixels in the same area. Coherence pixels are therefore spaced at approximately $3~\si{m}$ and $14~\si{m}$ in range and azimuth, respectively, however each coherence chip contains information from a $50~\si{m}$ by $70~\si{m}$ (15 by 5 SLC pixel) chip (Fig. \ref{fig:chip_stride}\textbf{(b)}). Note that the exact size of the chip on the ground depends on the local topography. The stride of the chip means that adjacent coherence chips in $\mathcal{D}_f$ will share a large number of SLC pixels, and thus have a high degree of correlation in their time series.

Note that as the area of interest lies within the training region a small fraction of the coherence time series in $\mathcal{D}_t$ will also be in $\mathcal{D}_f$, leading to a small amount of data leakage (i.e. the model is trained on some of the time series that it is trying to forecast). This could lead to some amount of overconfidence in the coherence forecast, reducing the quality of the damage map. However, model selection is done using the validation set ($\mathcal{D}_{t,v}$), i.e. the 20\% of $\mathcal{D}_t$ that is not used for training, and the data leakage is a tiny fraction of the total number of time series used for training.

\subsection*{August 24, 2016 \Mw{6.2} central Italy earthquake}
\begin{figure*}
    \centering
    \includegraphics[width=\textwidth]{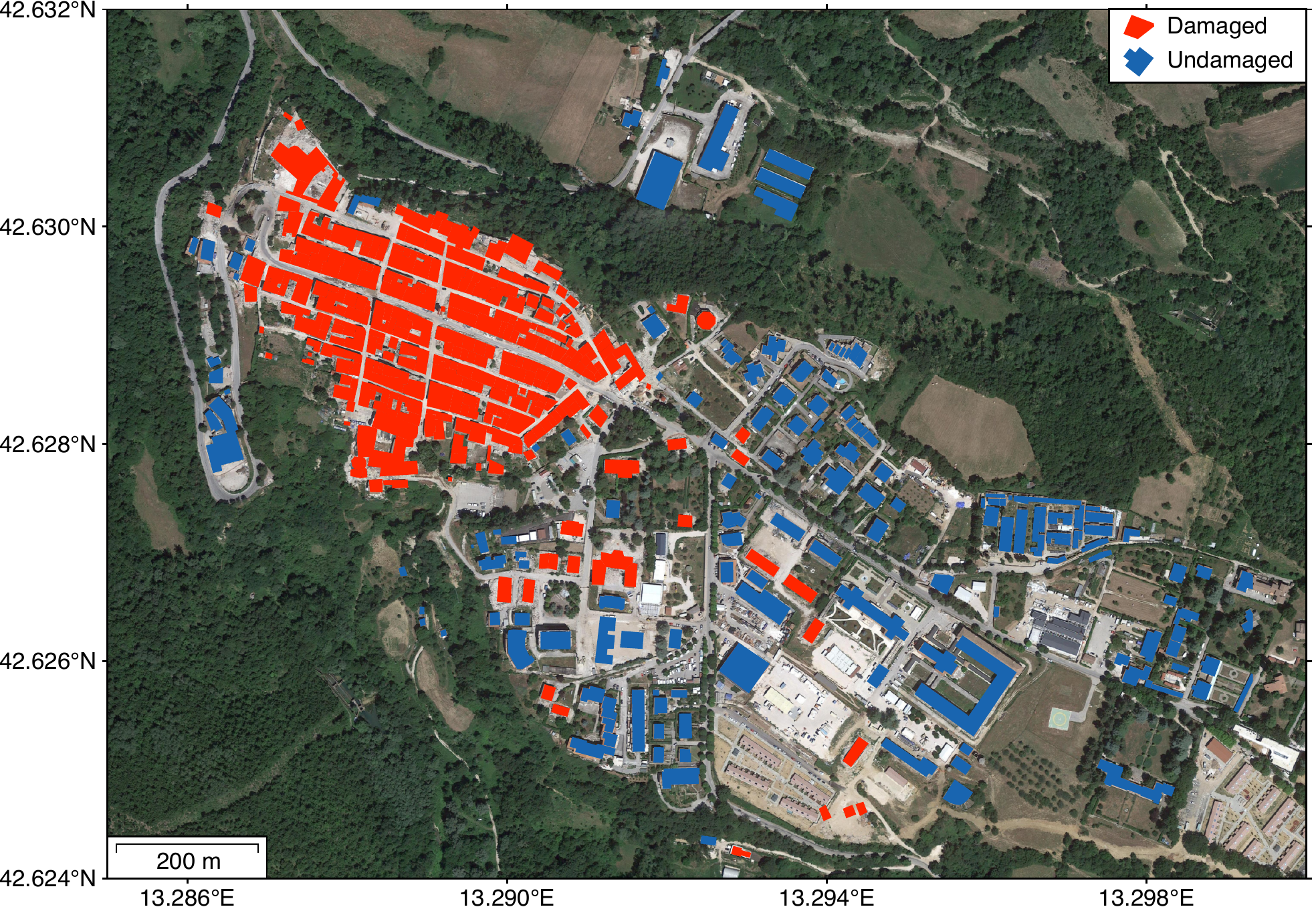}
    \caption{Ground truth data used for quantifying damage classification performance for the town of Amatrice, damaged during the August 24, 2016 \Mw 6.2 central Italy Earthquake. Building polygons and damage levels are supplied by the Copernicus Emergency Management Service  \cite{copernicus_emergency_management_service_emsr177_2016}, then simplified as described in Section \ref{a:study_areas}. Note that the optical imagery in this figure was taken a little under a year after the earthquake, and structures have been built that do not appear in the damage assessment. Optical imagery for the figure is from Google, taken July 6th 2017.}
    \label{fig:amatrice_ground}
\end{figure*}

Our first data set is from the \Mw{6.2} August 24, 2016 central Italy earthquake. 
At least 299 people were killed, with more than 400 injured and major damage done to the Italian town of Amatrice \cite{united_states_geological_survey_m_2016}. For this event, the Copernicus Emergency Management Service produced a damage map by manually reviewing high resolution pre- and post-event optical satellite imagery. Every building in Amatrice had its geographic footprint determined and each footprint was classified in to one of five damage levels, from ``unaffected'' to ``completely destroyed''. This data was made available online \cite{copernicus_emergency_management_service_emsr177_2016}. This data set was constructed via visual inspection of high resolution optical imagery taken before and after damage to the town. In our work we further simplify the data set by putting damage levels ``not affected'' and ``negligible to slight damage'' into an ``undamaged'' class and ``moderately damaged'', ``highly damaged'' and ``completely destroyed'' into a ``damaged'' class. The resulting damage map is plotted in Fig. \ref{fig:amatrice_ground}. As every building within the centre of the town was assessed, we can use this as ground truth with which to quantitatively compare the RNN and CCD methods.

The pre-event SAR repeat time was irregular, mostly 12 or 24 days between acquisitions, which creates a source of noise for our RNN method due to variable temporal decorrelation. The data in the training region, $\mathcal{D}_t$, covers the area around and east of Amatrice.

\subsection*{November 12, 2017 \Mw{7.3} Iran-Iraq earthquake}
\begin{figure*}
    \centering
    \includegraphics[width=\textwidth]{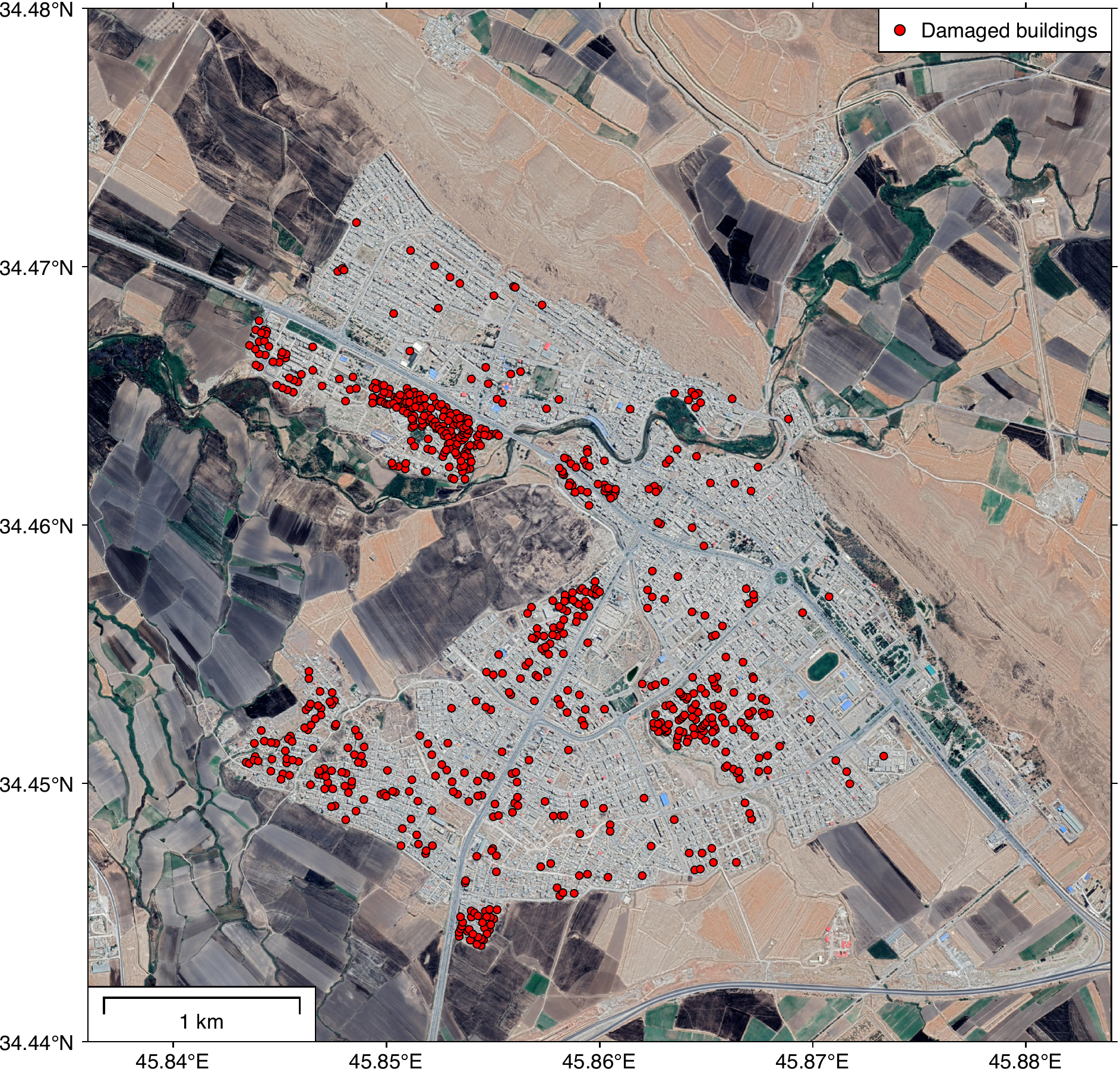}
    \caption{Location of 683 potentially damaged buildings manually mapped from optical satellite imagery by the United Nations Institute for Training and Research \cite{united_nations_institute_for_training_and_research_damaged_2017} for the town of Sarpol-e-Zahab, damaged during the November 12, 2017 \Mw{7.3} Iran-Iraq earthquake. Optical imagery for the figure is from Google, CNES/Airbus, taken July 27th 2020.}
    \label{fig:sarpol_ground}
\end{figure*}

The city of Sarpol-e-Zahab lies on the Iran-Iraq border and was hit by a \Mw{7.3} earthquake on November 12, 2017, causing substantial damage, over 7000 injured and at least 630 fatalities \cite{united_states_geological_survey_m_2017}. The United Nations Institute for Training and Research (UNITAR) produced an earthquake damage map for Sarpol-e-Zahab in the aftermath of the earthquake. By manually reviewing high resolution post-event optical satellite data, UNITAR mapped the location of 683 ``potentially damaged'' buildings \cite{united_nations_institute_for_training_and_research_damaged_2017} (Fig. \ref{fig:sarpol_ground}). However, the data does not contain damage levels or building footprints for every building, meaning we are unable to constrain which buildings are undamaged and so can't perform the same quantification as for Amatrice.

Before the earthquake, the satellite repeat time was 12 days (with six 24 day intervals). The post-event acquisition was six days after the final pre-event image. The data in the training region, $\mathcal{D}_t$, spans the city and the region to the north and west of Sarpol-e-Zahab.

\subsection*{July 2019 \Mw{6.4} and \Mw{7.1} Ridgecrest, California, USA earthquakes}
Numerous groups conducted mapping of surface ruptures, liquefaction and building damage in the aftermath of the Ridgecrest earthquakes \cite{stewart_preliminary_2019, hough_nearfield_2020,kendrick_geologic_2019, ponti_documentation_2020, zimmaro_liquefaction_2020}, allowing for a qualitative comparison with our damage proxy map (Fig. \ref{ext-fig:ridgecrest_dpms}).
Our area of interest is significantly larger than the previous two, and to cover the entire rupture we need to use the whole of $\mathcal{D}_t$ in $\mathcal{D}_f$. We use strides of 5 and 15 (the same as for $\mathcal{D}_t$) as we are more interested in larger scale signals. The fact that the training and forecasting regions are the same means that the model will have been trained on 80\% of $\mathcal{D}_f$, however as usual the optimum model is chosen from the performance on the 20\% of $\mathcal{D}_t$ ($\mathcal{D}_{t,v}$) that is not seen during training.  

The pre-event SAR acquisition interval in the Ridgecrest region was more variable than data for the other two areas, starting out at a modal value of 24 days in October 2014, then reducing to 12 days and finally 6 day intervals from March 2019 until the event. Again, the post-event acquisition was six days after the final pre-event image. While the acquisition interval was overall more variable than the other two regions, the several months of 6 day intervals before the event likely improves the quality of the RNN forecast. 

\begin{table*}
\caption{Summary of SAR data used for each study area. N\textsubscript{$aqn$}: number of pre-event acquisitions used, N\textsubscript{$t,f$}: number of pixels in the training ($\mathcal{D}_t$) and forecasting ($\mathcal{D}_f$) sets, respectively,  E\textsubscript{$t,f$}: spatial extent of the training and forecasting regions, Track: \mbox{Sentinel-1} orbital track and direction of flight.}
 \begin{tabularx}{\textwidth}{|X|X|X|X|X|X|X|X|}
 
\hline
 Event & Date (MM/DD/YY) & N\textsubscript{$aqn$} & N\textsubscript{$t$} & E\textsubscript{$t$} ($\si{km} \times\si{km}$) & N\textsubscript{$f$} & E\textsubscript{$f$} ($\si{km}\times\si{km}$) & Track \\
 \hline
 Central Italy & 08/24/16 & 48 & 3,396,828 & $140\times 70$ & 70,000 & $2.4\times 1.4$ & Descending 22 \\ 
 \hline
 Iran-Iraq & 11/12/17 & 89 & 11,056,389 & $210 \times 210$ & 750,000 & $5.2 \times 7.0$ & Ascending 174 \\ 
 \hline
 Ridgecrest & 07/04/19 & 97 & 2,860,000 & $60 \times 60$ & 2,860,000 & $60 \times 60$ & Ascending 64 \\ 
 \hline
\end{tabularx}
 \label{table:study_areas}
\end{table*}

\section{Training with a variable amount of pre-event data}\label{a:training_data}
In order to explore how the quality of the RNN damage map depends on the amount of data, we repeat the training and forecasting for the Amatrice data set, decreasing the temporal span of the pre-event data used and calculating the Precision-Recall Area Under Curve (PR AUC) for each trained model. We use the Amatrice data set for this exploration, as this is as this is the only location where we have comprehensive ground truth and so are able to obtain precision-recall curves. Training and forecasting data is generated as described in Section \ref{a:study_areas}, but then only coherence images that use data acquired on or after specifics dates are used in training. 

\begin{figure}
    \centering
    \includegraphics[width=0.47\textwidth]{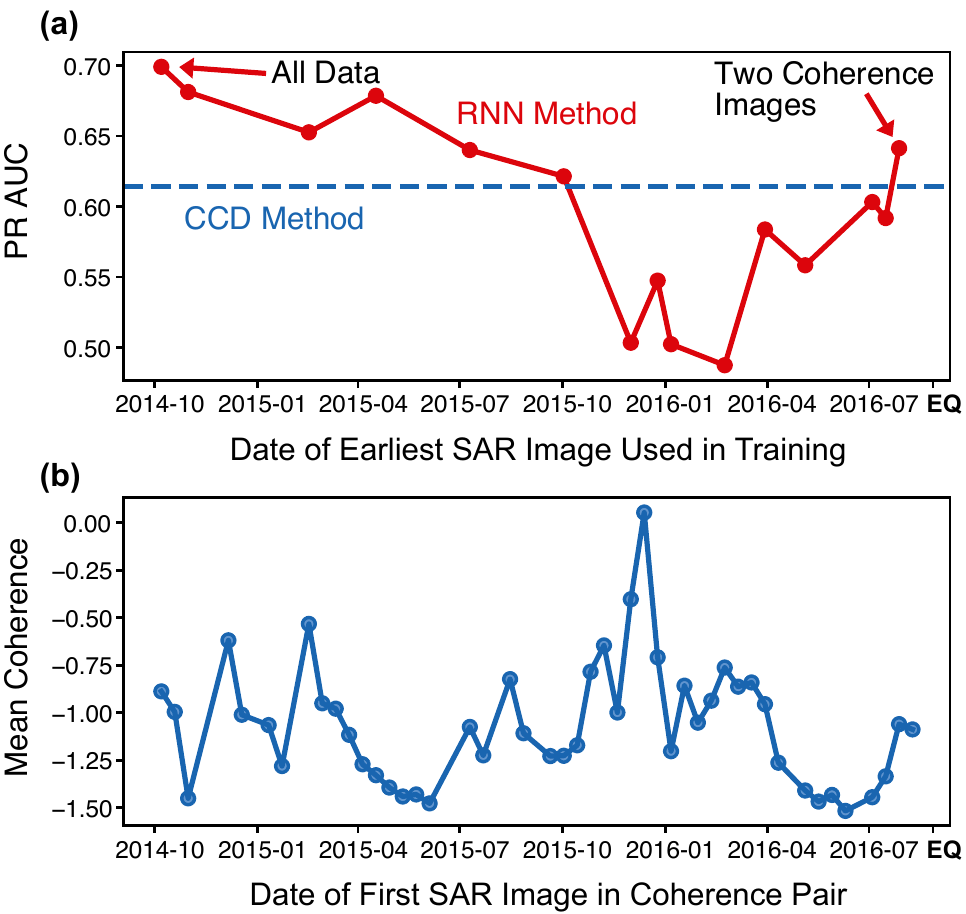}
    \caption{\textbf{(a)} RNN Precision-recall area under curve (PR AUC) for the Amatrice data set, with varying amounts of pre-event training and forecasting coherence data (red line). Only data acquired on or after the indicated date is used in training and forecasting. The PR AUC for the CCD method is also presented for comparison (horizontal dashed blue line). Training that uses all data is indicated at the top left. Training that only uses two pre-event coherence images is indicated on the right hand side.  
   \textbf{(b)} Mean of the logit transform of the squared coherence against the date of the first SAR acquisition in that coherence image. Values are presented up to the final pre-event coherence image. The co-event coherence is not used in training, so its mean is not presented here. Generally SAR images in each coherence image are separated by 12 or 24 days.
   The date of the August 2016 central Italy earthquake is indicated by "\textbf{EQ}" on the right of the both plots.}
    \label{fig:pre_num_pr_auc}
\end{figure}

These results are presented in Figure \ref{fig:pre_num_pr_auc}\textbf{(a)}, along with a comparison to the CCD PR AUC as a baseline. The CCD method uses a single pre-event coherence image along with the co-event image. When we use all available data, we obtain the result previously presented in Figure \ref{ext-fig:amatrice_pr_f05} (PR AUC=0.7). As we move the cut-off date later in time, we observe a steady decrease in the PR AUC until late 2015, where there is a sudden drop to around 0.5 (well below CCD performance), followed by an overall increase up to the final point, which only uses two pre-event coherence images.

If every coherence image was adding information that could allow for a better forecast of the coherence distribution, then we would expect that more training data would improve the results. The sudden drop in PR AUC when the data cut-off is in late 2015, followed by the increase in performance as we reduce the amount of training data, conflicts with these expectations. This behavior suggests that the coherence data in the year leading up to the earthquake are potentially less representative of the distribution that we are trying to forecast, meaning that training only on this data decreases performance. 

We can gain some indication of the variability of coherence through time by looking at the mean of coherence data (in logit space) used for training, presented in Figure \ref{fig:pre_num_pr_auc}\textbf{(b)}. These results show anomalously high coherence values in late 2015, lining up with the decrease in PR AUC. It's important to note that all training with a cut-off date earlier than late 2015 includes these anomalously high coherence values. However, it is possibly the case that as we increase the amount of data beyond a year, the fraction of the training data which is anomalous decreases and so the results improve.

The results presented in this section suggest that performance can be unreliable when training is done on less than a year of coherence data. However this conclusion is likely highly dependent on the specific attributes of the local region. For example, we would probably not observe the same effects in a dry desert area, where coherence is more stable through time. Further testing is necessary to explore how the results depend on the time span of the training data.

As noted in Section \ref{a:rnn_details}, non-determinism in the RNN training leads to variability in the exact PR AUC values when training is repeated for the same data. This variability leads to some scatter in the results, however it does not affect the overall results presented in this section. 

\bibliographystyle{IEEEtran}
\bibliography{main}